\documentclass[a4paper, 11pt]{article}
\usepackage[pdftex, a4paper]{geometry}

\usepackage{amsfonts}
\usepackage{dsfont}
\usepackage{mathrsfs}
\usepackage{amssymb}
\usepackage{bbm}
\usepackage{epsfig}
\usepackage{amsmath}
\usepackage{appendix}
\usepackage[english]{babel}
\usepackage[all]{xy}

\usepackage{fancyhdr}
\newtheorem{them}{Theorem}
\newtheorem{defn}{Definition}
\newtheorem{lem}{Lemma}
\newtheorem{pro}{Proposition}
\newtheorem{remark}{Remark}
\newtheorem{assum}{Assumption}

\def\d{ {\rm d  }}
\def\sgn{{\rm sgn}}

\def\th{\mbox{\small th}}
\def\T{^{\rm\tiny T}}

\def\Int{{\rm int}}
\def\co{{\rm co}}

\def\cs{{\rm cs}}
\def\ri{{\rm ri}}
\def\rb{{\rm rb}}

\begin{document}

\title{\bf Multi-agent Systems with Compasses}
\date{}

\author{Ziyang Meng,
Guodong Shi, and Karl Henrik Johansson\thanks{A brief version of this work was presented in the 33rd Chinese Control Conference, Nanjing, China, July 2014 \cite{MengZiyang_CCC2014}. Z. Meng is with Department of Precision Instrument, Tsinghua University, Beijing 100084, China.
G. Shi is with College of Engineering and Computer Science, The Australian National University, Canberra, Australia.
K. H. Johansson is with ACCESS Linnaeus Centre, School of Electrical Engineering,
Royal Institute of Technology, Stockholm, Sweden.
This work has been supported in part
by the Knut and Alice Wallenberg Foundation, the Swedish Research
Council, and the Alexander von Humboldt Foundation of Germany. Email: { {\tt {ziyang.meng@tum.de, guodong.shi@anu.edu.au, kallej@ee.kth.se}}.}} }
\maketitle

\begin{abstract}
This paper investigates agreement protocols over cooperative and cooperative--antagonistic multi-agent networks with  coupled continuous-time nonlinear dynamics. To guarantee convergence for such systems, it is common in the literature to assume that the vector field of each agent is pointing inside the convex hull formed by the states of the agent and its neighbors, given that the relative states between each agent and its neighbors are available. This convexity condition is relaxed in this paper, as we show that it is enough that the vector field belongs to a strict tangent cone based on a local supporting hyperrectangle. The new condition has the natural physical interpretation of requiring shared reference directions in addition to the available local relative states. Such shared reference directions can be further interpreted as if each agent holds a magnetic compass indicating the orientations of a global frame. It is proven that the cooperative multi-agent system achieves exponential state agreement if and only if the time-varying interaction graph is uniformly jointly quasi-strongly connected. Cooperative--antagonistic multi-agent systems are also considered. For these systems, the relation has a negative sign for arcs corresponding to antagonistic interactions. State agreement may not be achieved, but instead it is shown that all the agents' states asymptotically converge, and their limits agree componentwise in absolute values if and in general only if the time-varying interaction graph is uniformly jointly strongly connected.
\end{abstract}


\section{Introduction}

In the last decade, coordinated control of multi-agent systems has attracted extensive attention due to its broad applications in engineering, physics, biology and social sciences, e.g., \cite{CaoMing_TAC08,JadbabaieLinMorse03,Baras_SIAM13,SaberFaxMurray07_IEEE,VicsekEtAl95}. A fundamental idea is that by carefully implementing distributed control protocols for each agent, collective tasks can be reached for the overall system using only neighboring information exchange. Several important results have been established, e.g., in the area of mobile systems including spacecraft formation flying, rendezvous of multiple robots, and animal flocking \cite{LinBrouckeFrancis04,Cortes_TAC2006,TannerJadbabaiePappas07}.

Agreement protocols, where the goal is to drive the states of the agents to reach a common value using local interactions, play a basic role in coordination of multi-agent systems. The state agreement protocol and its fundamental convergence properties were established for linear systems in the classical work \cite{Tsitsiklis_TAC86}.
The convergence of the linear agreement protocol has been widely studied since then for both continuous-time and discrete-time models, e.g., \cite{Blondel_CDC2005,JadbabaieLinMorse03,RenBeard05_TAC}. Much understandings have been established, such as the explicit convergence rate in many cases \cite{CaoMing_SIAM08J2,Moreau_CDC04,SaberMurray04,Olshevsky_SIAM09}.
A major challenge is how to quantitatively characterize the influence of a time-varying communication graph on the agreement convergence. Agreement protocols with nonlinear dynamics have also drawn attention in the literature, e.g., \cite{Bauso_SCL2006,HendrickxTAC13,LinZhiyun_SIAM07,Moreau_TAC05,ShiGuodong_Automatica2009,ShiSIAM}. Due to the complexity of nonlinear dynamics, it is in general difficult to obtain explicit convergence rates for these systems. All the above studies on linear or nonlinear multi-agent dynamics are based on the standing assumption that agents in the network are cooperative. Recently,
motivated from opinion dynamics evolving over social networks \cite{Easley_Book,Wasserman_Book}, state agreement problems over cooperative--antagonistic networks were introduced \cite{AltafiniPlosOne,Altafini_TAC13}. In such networks, antagonistic neighbors exchange their states with opposite signs compared to cooperative neighbors.

In most of the work discussed above, a convexity assumption plays an essential role in the local interaction rule for reaching state agreement. For discrete-time models, it is usually assumed that each agent updates its state as a convex combination of its neighbors' states \cite{Blondel_CDC2005,JadbabaieLinMorse03}. A precise characterization of this convexity condition guaranteeing asymptotic agreement was established in \cite{Moreau_TAC05}. For continuous-time models, an interpretation of this assumption is that the vector field for each agent must fall into the relative interior of a tangent cone formed by the convex hull of the relative state vectors in its neighborhood \cite{LinZhiyun_SIAM07}. The recent work \cite{Baras_TAC13} generalized agreement protocols to convex metric spaces, but a convexity assumption for the local dynamics continued to play an important role in ensuring agreement convergence.

In this paper, we show that the convexity condition for agreement seeking of multi-agent systems can be relaxed at the cost of shared reference directions.
Such shared reference directions can be easily obtained by a magnetic compass, with the help of which the direction of each axis can be observed from a prescribed global coordinate system.
Using the relative state information and the shared reference direction information, each agent can derive a strict tangent cone from a local supporting hyperrectangle. This cone defines the feasible set of local control actions for each agent to guarantee convergence to state agreement. In fact, the agents just need to
determine, through sensing or communication, the relative orthant of each of their neighbors' states. The vector field of an agent can be outside of the convex hull formed by the states of the
agent and its neighbors, so this new condition provides a relaxed condition for agreement seeking. We remark that a compass is naturally present in many systems. For instance, the classical Vicsek's model \cite{VicsekEtAl95} inherently uses ``compass''-like directional information and the calculation of each agent's heading relies on the information where the common east is. In addition, scientists observed that
the European Robin bird can detect and navigate through the Earth's magnetic field, providing them with biological compasses in addition to their normal vision \cite{Compass_Biophysical12}. Engineering systems, such as multi-robot networks, can be equipped with magnetic compasses at a low cost \cite{Compass-Ref2,Compass-Ref1}.

Under a general definition of nonlinear multi-agent systems with shared reference directions, we establish two main results:
\begin{itemize}
\item  For cooperative networks, we show that the underlying graph associated with the nonlinear interactions being uniformly jointly quasi-strongly connected is necessary and sufficient for exponential agreement. The convergence rate is explicitly given. This improves the existing results based on convex hull conditions \cite{Moreau_TAC05,LinZhiyun_SIAM07}.

\item For cooperative-antagonistic networks, we propose a general model following the sign-flipping interpretation along an antagonistic arc introduced in \cite{Altafini_TAC13}. We show that when the underlying graph is uniformly jointly strongly connected, irrespective with the sign of the arcs, all the agents' states asymptotically converge, and their limits agree componentwise in absolute values.

\end{itemize}

The remainder of the paper is organized as follows. In
Section \ref{sec:prim}, we give some mathematical preliminaries on convex sets, graph theory, and Dini derivatives.
The nonlinear multi-agent dynamics, the interaction graph, the shared reference direction, and the agreement metrics are given in Section \ref{sec:ps}.
The main results and discussions are presented in Section \ref{sec:results}. The proofs of the results are presented in Sections \ref{sec:cooperative} and \ref{sec:antagonistic}, respectively, for cooperative and cooperative--antagonistic networks. A brief concluding remark is given in
Section \ref{sec:conclusion}.

\section{Preliminaries}\label{sec:prim}
In this section, we introduce some mathematical preliminaries on convex analysis \cite{Aubin1991}, graph theory \cite{Godsil_Book}, and Dini derivatives \cite{Filippov_Book}.

\subsection{Convex analysis}\label{sec:convex}
For any nonempty set $\mathcal{S} \subseteq \mathbb{R}^d$, $\|x\|_{\mathcal{S}}=\inf_{y\in \mathcal{S}}\|x-y\|$ is called  the distance between $x\in \mathbb{R}^d$ and $\mathcal{S}$, where $\|\cdot\|$ denotes the Euclidean norm. A set $\mathcal{S}\subset \mathbb{R}^d$ is called convex if $(1-\zeta)x+\zeta y\in \mathcal{S}$ when $x\in\mathcal{S}$, $y\in\mathcal{S}$, and $0\leq \zeta\leq 1$.  A convex set ${\mathcal{S}}\subset \mathbb{R}^d$ is called a convex cone if $\zeta x\in {\mathcal{S}}$ when $x\in {\mathcal{S}}$ and $\zeta>0$. The convex hull of $\mathcal{S}\subset \mathbb{R}^d$ is denoted $\co(\mathcal{S})$ and the convex hull of a finite set of points $x_1,x_2,\dots,x_n\in \mathbb{R}^d$ denoted $\co\{x_1,x_2,\dots,x_n\}$.

Let $\mathcal{S}$ be a convex set. Then there is a unique element $P_{\mathcal{S}}(x)\in\mathcal{S}$, called the convex projection of $x$ onto $\mathcal{S}$,  satisfying $\|x-P_{\mathcal{S}}(x)\|=\|x\|_{\mathcal{S}}$ associated to any $x\in \mathbb{R}^d$. It is also known that $\|x\|_{\mathcal{S}}^2$ is continuously differentiable for all $x\in \mathbb{R}^d$, and its gradient can be explicitly computed \cite{Aubin1991}:
\begin{align}
\nabla\|x\|_{\mathcal{S}}^2=2(x-P_{\mathcal{S}}(x))\label{eq:set-deri}.
\end{align}
Let $\mathcal{S}\subset \mathbb{R}^d$ be convex and closed. The interior and boundary of $\mathcal{S}$ is denoted by $\Int(\mathcal{S})$ and $\partial\mathcal{S}$, respectively. If $\mathcal{S}$ contains the origin, the smallest subspace containing $\mathcal{S}$ is the carrier subspace denoted by $\cs(\mathcal{S})$. The relative interior of $\mathcal{S}$, denoted by $\ri(\mathcal{S})$, is the interior of $\mathcal{S}$ with respect to the subspace $\cs(\mathcal{S})$ and the relative topology used. If $\mathcal{S}$ does not contain the origin, $\cs(\mathcal{S})$ denotes the smallest subspace containing $\mathcal{S}-z$, where $z$ is any point in $\mathcal{S}$. Then, $\ri(\mathcal{S})$ is the interior of $\mathcal{S}$ with respect to the subspace $z+\cs(\mathcal{S})$. Similarly, we can define the relative boundary $\rb(\mathcal{S})$.

Let $\mathcal{S}\subset \mathbb{R}^d$ be a closed convex set and $x\in \mathcal{S}$. The tangent cone to $\mathcal{S}$ at $x$ is defined as the set
$
\mathcal{T}(x,\mathcal{S})=\Big\{z\in \mathbb{R}^d: \liminf_{\zeta\rightarrow 0}\frac{\|x+\zeta z\|_{\mathcal{S}}}{\zeta}=0\Big\}.
$
Note that if $x\in \Int(\mathcal{S})$, then $\mathcal{T}(x,\mathcal{S})=\mathbb{R}^d$. Therefore, the definition of $\mathcal{T}(x,\mathcal{S})$ is essential only when $x\in \partial\mathcal{S}$.
The following lemma can be found in \cite{Aubin1991} and will be used.
\begin{lem}\label{lem:cone}
Let $\mathcal{S}_1,\mathcal{S}_2\subset \mathbb{R}^d$ be convex sets. If $x\in\mathcal{S}_1\subset\mathcal{S}_2$, then $\mathcal{T}(x,\mathcal{S}_1)\subset\mathcal{T}(x,\mathcal{S}_2)$.
\end{lem}

\subsection{Graph theory}\label{sec:graph}

A directed graph $\mathcal{G}$ consists of
a pair $(\mathcal{V}, \mathcal{E})$, where $
\mathcal{V}=\{1,2,\ldots, {n}\}$ is a finite, nonempty set of nodes and
$ \mathcal{E} \subseteq  \mathcal{V}\times \mathcal{V}$ is a set of
ordered pairs of nodes, denoted arcs. The set of neighbors of node $ i$ is denoted $\mathcal{N}_i :=
\{j:( j, i)\in \mathcal{E}\}$.
A directed path in a directed graph is a sequence of arcs of the form $(i,j),(j,k),\dots$. If there exists a path from node $i$ to $j$, then node $j$ is said to be reachable from node $i$. If for node $  i$, there exists a path from $  i$ to any other node, then $ i$ is called a root of $\mathcal{G}$. $\mathcal{G}$ is said to be strongly connected if each node is reachable from any other node. $\mathcal{G}$ is said to be quasi-strongly connected if $\mathcal{G}$ has a root.

\subsection{Dini derivatives}

Let $D^+V(t,x(t))$ be the upper Dini derivative of $V(t,x(t))$ with respect to $t$, i.e.,
$
D^+V(t,x)=\limsup_{\tau\rightarrow 0^+}\frac{V(t+\tau,x(t+\tau))-V(t,x(t))}{\tau}.
$
The following lemma \cite{Danskin66} will be used for our analysis.
\begin{lem}\label{lem:Dini}
Suppose for each $i\in \mathcal{V}$, $V_i:\mathbb{R}\times \mathcal{M}\rightarrow \mathbb{R}$ is continuously differentiable. Let $V(t,x)=\max_{i\in \mathcal{V}} V_i(t,x)$, and let $\widehat{\mathcal{V}}(t)=\{i\in \mathcal{V}: V_i(t,x(t))=V(t,x(t))\}$ be the set of indices where the maximum is reached at time $t$. Then
$
D^+V(t,x(t))=\max_{i\in \widehat{\mathcal{V}}(t)}\dot V_i(t,x(t)).
$
\end{lem}

\section{Multi-agent Network Model}\label{sec:ps}
In this section, we present the model of the considered multi-agent systems, introduce the corresponding interaction graph, and define some useful geometric concepts used in the control laws.

Consider a multi-agent system with agent set $\mathcal{V}=\{1,\dots,n\}$. Let $x_i\in\mathbb{R}^d$  denote the state of agent $i$. Let $x=(x_1\T, x_2\T,\dots, x_n\T)\T$ and denote $\mathcal{D}=\{1,2,\dots,d\}$.

\subsection{Nonlinear multi-agent dynamics}\label{sec:nonlinear dynamics}
Let $\mathfrak{P}$ be a given (finite or infinite) set  of indices. An element in $\mathfrak{P}$ is denoted by $p$. For any $p\in \mathfrak{P}$, we define a function
$
f_p(x_1,x_2,\dots,x_n): \mathbb{R}^{dn}\rightarrow \mathbb{R}^{dn}
$
associated with $p$, where $
f_p(x_1,x_2,\dots,x_n)=\left( \begin{array}{ccc}
f_p^1(x_1,x_2,\dots,x_n) \\
\vdots  \\
f_p^n(x_1,x_2,\dots,x_n) \end{array} \right)
$
with $f_p^i:\mathbb{R}^{dn}\rightarrow \mathbb{R}^{d}$, $i=1,2,\dots,n$.

Let $\sigma(t):[t_0,\infty)\rightarrow \mathfrak{P}$ be a piecewise constant function, so, there exists a sequence of increasing time instances $\{t_l\}_0^\infty$ such that $\sigma(t)$ remains constant for
$t\in[t_{l},t_{l+1})$  and switches at $t=t_{l}$.

The dynamics of the multi-agent systems is described by the switched nonlinear system
  \begin{equation}
\dot x(t)=f_{\sigma(t)}(x(t)).\label{eq:switch}
  \end{equation}
We place some mild assumptions on this system.

\begin{assum}\label{assm0}
There exists a lower bound $\tau_d>0$, such that $\inf_{l}(t_{l+1}-t_l)\geq \tau_d$.
\end{assum}

\begin{assum}\label{assm1}
$f_p(x)$ is uniformly locally Lipschitz on $\mathbb{R}^{dn}$, i.e., for every $x\in \mathbb{R}^{dn}$, we can find a neighborhood  $\mathcal{U}_\alpha(x)=\{y\in \mathbb{R}^{dn}: \|y-x\|\leq \alpha\}$ for some $\alpha>0$ such that there exists a real number $L(x)>0$ with $\|f_p(a)-f_p(b)\|\leq L(x)\|a-b\|$ for any $a,b\in \mathcal{U}_\alpha(x)$ and $p\in \mathfrak{P}$.
\end{assum}

Under Assumptions \ref{assm0} and \ref{assm1}, the Caratheodory solutions of (\ref{eq:switch}) exist
for arbitrary initial conditions, and they are absolutely continuous functions  for almost all $t$ on the maximum interval of existence \cite{Filippov_Book}. All our further discussions will be on the Caratheodory solutions of (\ref{eq:switch}) without specific mention.

\subsection{Interaction graph}\label{sec:interaction graph}
Having the dynamics defined for the considered multi-agent system, similar to \cite{LinZhiyun_SIAM07}, we introduce next its interaction graph.

\begin{defn} The graph $\mathcal{G}_p=(\mathcal{V}, \mathcal{E}_p)$ associated with $f_p$ is the directed graph on node set $\mathcal{V}=\{ 1, 2,\dots, n\}$  and arc set $\mathcal{E}_p$ such that  $( j, i)\in \mathcal{E}_p$ if and only if $f_p^i$ depends on $x_j$, i.e., there exist $x_j,\overline{x}_j\in \mathbb{R}^d$ such that
$
f_p^i(x_1,\dots,x_j,\dots,x_n)\neq f_p^i(x_1,\dots,\overline{x}_j,\dots,x_n).
$
\end{defn}

The set of neighbors of node $i$ in $\mathcal{G}_p$ is denoted by $\mathcal{N}_i(p)$. The dynamic interaction graph associated with system \eqref{eq:switch} is denoted by $\mathcal{G}_{\sigma(t)}=(\mathcal{V}, \mathcal{E}_{\sigma(t)})$. The joint graph of $\mathcal{G}_{\sigma(t)}$ during time interval $[t_1,t_2)$ is defined by $\mathcal{G}_{\sigma(t)}([t_1,t_2))=\bigcup_{t\in[t_1,t_2)}\mathcal{G}(t)=(\mathcal{V},
\bigcup_{t\in[t_1,t_2)}\mathcal{E}_{\sigma(t)})$. We impose the following definition  on the connectivity of $\mathcal{G}_{\sigma(t)}$, cf., \cite{ShiSIAM}.
\begin{defn}
$\mathcal{G}_{\sigma(t)}$ is uniformly jointly quasi-strongly (respectively, strongly) connected if there exists a constant $T>0$ such that $\mathcal{G}([t,t+T))$ is quasi-strongly (respectively, strongly) connected for any $t\geq t_0$.
\end{defn}

For each $p\in\mathfrak{P}$, the node relation along an interaction arc $(j,i)\in \mathcal{E}_p$ may be cooperative, or antagonistic. We assume that there is a sign, ``+1'' or ``-1'', associated with each $(j,i)\in \mathcal{E}_p$, denoted by $\sgn^{ij}_p$. To be precise, if $j$ is cooperative to $i$, $\sgn^{ij}_p=+1$, and if $j$ is antagonistic to $i$, $\sgn^{ij}_p=-1$.

\begin{defn}
If $\sgn^{ij}_p=+1$, for all $(j, i)\in \mathcal{E}_p$ and all $p\in\mathfrak{P}$, the considered multi-agent network is called a cooperative network. Otherwise, it is called a cooperative-antagonistic network.
\end{defn}

\subsection{Shared reference direction, hyperrectangle, and tangent cone} \label{sec:hypercube}

We assume that each agent has access to shared reference directions with respect to a common Cartesian coordinate system. We use $(\overrightarrow{r_1},\overrightarrow{r_2},\dots,\overrightarrow{r_d})$ to represent the basis of that $\mathbb{R}^d$ Cartesian coordinate system.  Here $\overrightarrow{r_k}=(0,\dots,0,1,0,\dots,0)$ denotes the unit vector in the direction of axis $k\in \mathcal{D}$.

A hyperrectangle is the generalization of a rectangle to higher dimensions. An axis-aligned hyperrectangle is a hyperrectangle subject to the constraint that the edges of the hyperrectangle are parallel to the Cartesian coordinate axes.

\begin{defn}\label{def:cube}
Let $\mathcal{C}\subset \mathbb{R}^d$ be a bounded set.
The supporting hyperrectangle $\mathcal{H}(\mathcal{C})$ is the axis-aligned hyperrectangle
$
\mathcal{H}(\mathcal{C})=[\min(\mathcal{C})_1,\max(\mathcal{C})_1]\times[\min(\mathcal{C})_2,\max(\mathcal{C})_2]
\times\dots\times[\min(\mathcal{C})_d,\max(\mathcal{C})_d],
$
where by definition $\min(\mathcal{C})_k:=\min_{ y\in \mathcal{C}}y_k$, $\max(\mathcal{C})_k:=\max_{ y\in \mathcal{C}}y_k$, and $y_k$ denotes the $k$th entry of $y$.
\end{defn}

In other words, a supporting hyperrectangle of a bounded set $\mathcal{C}$ is an axis-aligned minimum bounding hyperrectangle.

\begin{figure}
\begin{center}
\includegraphics[scale=0.25]{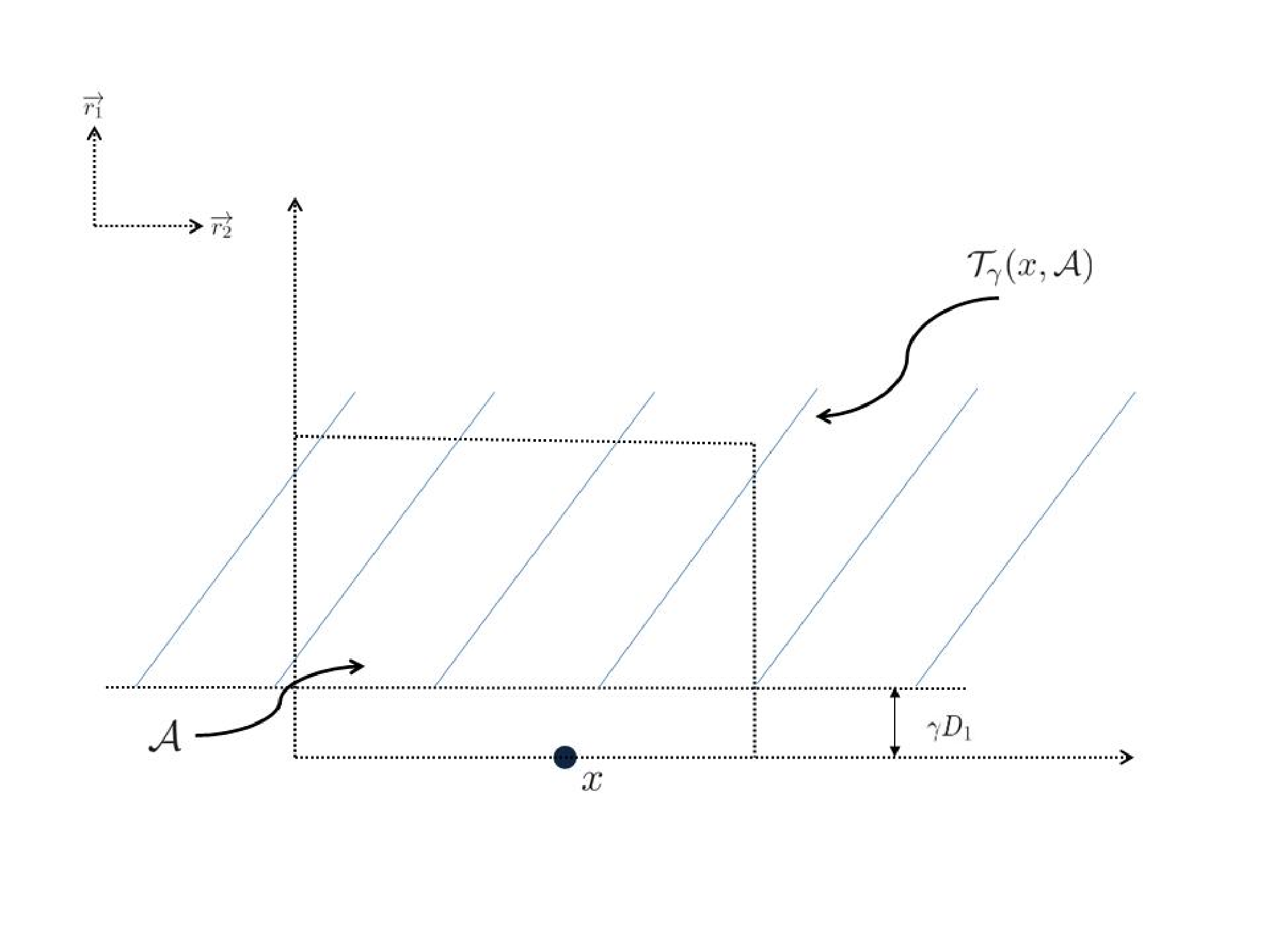}
\caption{An example of the $\gamma$-strict tangent cone. }\label{fig-gamma-cone}
 \end{center}
\end{figure}

\begin{defn}\label{def:cone}
Let $\mathcal{A}\subset \mathbb{R}^d$ be an axis-aligned hyperrectangle and $\gamma>0$ a constant. The $\gamma$-strict tangent cone to $\mathcal{A}$ at $x\in \mathbb{R}^d$ is the set
\begin{equation}
\mathcal{T}_\gamma(x,\mathcal{A})=\begin{cases}
\cs(\mathcal{A}); \quad \text{if}~~ x\in \ri(\mathcal{A}) \\
\mathcal{T}(x,\mathcal{A})\bigcap_{k\in\mathcal{I}}\{z\in\mathbb{R}^d:\left|\langle z,\overrightarrow{r_k}\rangle\right|\geq \gamma D_k(\mathcal{A})\};\quad \text{otherwise},
\end{cases}
\end{equation}
where $\mathcal{I}=\{k\in\mathcal{D}:x\in\rb_{k}(\mathcal{A}) \}$, $\rb_{k}(\mathcal{A})$ denotes the two facets of $\mathcal{A}$ perpendicular to the axis $\overrightarrow{r_k}$, and $D_k(\mathcal{A})=\max(\mathcal{A})_k-\min(\mathcal{A})_k$ denotes the side length parallel to the axis $\overrightarrow{r_k}$.
\end{defn}
Figure~\ref{fig-gamma-cone} gives an example of the $\gamma$-strict tangent cone to $\mathcal{A}$ at $x$.

\subsection{Agreement metrics}

We next define uniformly asymptotic agreement and exponential agreement in this section.

\begin{defn}\label{def:asymptotic}
The multi-agent system \eqref{eq:switch} is said to achieve uniformly asymptotic agreement on $\mathcal{S}_0\subseteq \mathbb{R}^d$ if

\begin{enumerate}
\item[(i).] point-wise uniform agreement can be achieved, i.e., for all $\eta\in\mathcal{J}$,  and $\varepsilon>0$, there exists $\delta(\varepsilon)>0$ such that for all $ t_0\geq 0$,
$
\|x(t_0)-\eta\|< \delta \quad \Rightarrow \quad \|x(t)-\eta\|<\varepsilon,~~ \forall t\geq t_0,
$
where $x(t_0)\in \mathcal{S}_0^n$, and the agreement manifold is defined as $\mathcal{J}=\{x\in \mathcal{S}_0^n:x_1=x_2=\dots=x_n\}$ and $\mathcal{S}_0^n$ denotes $\mathcal{S}_0\times\mathcal{S}_0\times\dots\mathcal{S}_0$; and

\item[(ii).] uniform agreement attraction can be achieved, i.e., for all
$\varepsilon>0$, there exist $\eta(x(t_0))\in\mathcal{J}$ and $T(\varepsilon)>0$ such that for all $t_0\geq 0$,
$
\|x(t)-\eta\|<\varepsilon, ~~\forall t\geq t_0+T.
$
\end{enumerate}
\end{defn}

\begin{defn}\label{def:exponential}
The multi-agent system \eqref{eq:switch} is said to achieve exponential state agreement on $\mathcal{S}_0\subseteq \mathbb{R}^d$ if
\begin{enumerate}
\item[(i).] point-wise uniform agreement can be achieved; and

\item[(ii).] exponential agreement attraction can be achieved, i.e.,
there exist $\eta (x(t_0))\in\mathcal{J}$ and $k(\mathcal{S}_0)>0$, $\lambda(\mathcal{S}_0)>0$, such that for all $t_0\geq 0$,
$
\|x(t)-\eta\|\leq ke^{-\lambda(t-t_0)}\|x(t_0)-\eta\|.
$
\end{enumerate}
\end{defn}

\section{Main Results}\label{sec:results}

In this section, we state the main results of the paper.

\subsection{Cooperative networks}

We first study the convergence property of the nonlinear switched system \eqref{eq:switch} over a cooperative network defined by an interaction graph. Introduce the local convex hull $\mathcal{C}_p^i(x)=\co\{x_i,x_j:j\in \mathcal{N}_i(p)\}$.
In order to achieve exponential agreement,
we propose the following strict tangent cone condition for the feasible vector field.
\begin{assum}\label{assm-strict}
For all $i\in \mathcal{V}$, $p\in \mathfrak{P}$, and $x\in\mathbb{R}^{dn}$, it holds that $f_p^i(x)\in\mathcal{T}_\gamma(x_i,\mathcal{H}(\mathcal{C}_p^i(x)))$.
\end{assum}

\begin{figure}
\begin{center}
\includegraphics[scale=0.48]{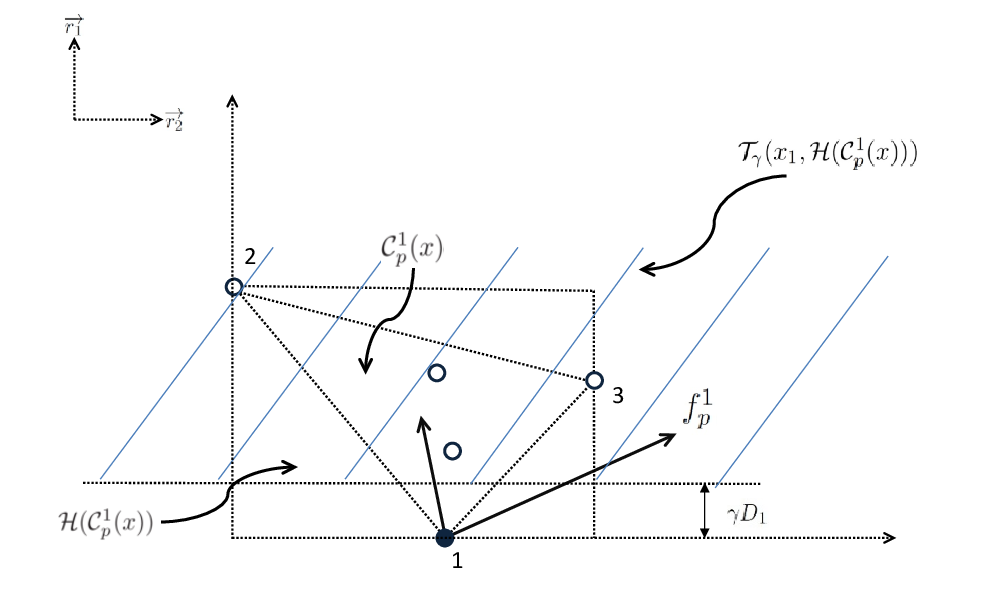}
\caption{Convex hull, supporting hyperrectangle, strict tangent cone and feasible vectors $f_p^i$ satisfying Assumption \ref{assm-strict}.}\label{fig2}
 \end{center}
\end{figure}

In Assumption \ref{assm-strict}, the vector $f_p^i$ can be chosen freely from the set $\mathcal{T}_\gamma(x_i,\mathcal{H}(\mathcal{C}_p^i(x)))$. Hence, the assumption specifies constraints on the feasible controls for the multi-agent system. Here $\mathcal{C}_p^i(x)$ denotes the convex hull formed by agent $i$ and its neighbors, $\mathcal{H}(\mathcal{C}_p^i(x))$  denotes the local supporting hyperrectangle of the set $\mathcal{C}_p^i(x)$, and $\mathcal{T}_\gamma(x_i,\mathcal{H}(\mathcal{C}_p^i(x)))$ denotes the $\gamma-$strict tangent cone to $\mathcal{H}(\mathcal{C}_p^i(x))$ at $x_i$.
Figure~\ref{fig2} gives an example of the convex hull and the supporting hyperrectangle formed by agent $1$ and its' neighbors. Two feasible vectors $f_p^1$ are also presented.

In order to implement a controller compatible with Assumption \ref{assm-strict}, the agents need to
determine, through local sensing or communication, the relative orthant of each of their neighbors' states.
This can be realized, for instance,  if each agent is capable of measuring the relative states with respect to its neighbors and is aware of the direction of each axis of a prescribed global coordinate system. More specifically, when the agent is in the interior of the hyperrectangle, the vector field for the agent can be chosen arbitrarily. When the agent is on the boundary of its supporting hyperrectangle, the feasible control is any direction pointing inside the tangent cone of its supporting hyperrectangle. Note that the absolute state of the agents is not needed, but each agent needs to identify $d-1$ absolute directions such that it can identify the direction of its neighbors with respect to itself.
For example, for the planar case $d=2$, in addition to the relative state measurements with respect to its  neighbors, each agent just needs to be equipped with a compass. The compass together with relative state measurements provide the quadrant location information of the neighbors.

We state an exponential agreement result for the cooperative multi-agent systems.
\begin{them}\label{thm1}
Suppose $\mathcal{S}_0$ is compact and that Assumptions \ref{assm0}, \ref{assm1}, and \ref{assm-strict} hold. Then, the cooperative multi-agent system \eqref{eq:switch} achieves exponential agreement on $\mathcal{S}_0$ if and only if its interaction graph $\mathcal{G}_{\sigma(t)}$ is uniformly jointly quasi-strongly connected.
\end{them}

In order to compare  the proposed ``supporting hyperrectangle condition'' with respect to the usual convex hull condition \cite{Moreau_TAC05,LinZhiyun_SIAM07}, we introduce the following assumption, which is a weaker condition than Assumption \ref{assm-strict}.

\begin{assum}\label{assm-hyper}
For all $i\in \mathcal{V}$, $p\in \mathfrak{P}$, and $x\in\mathbb{R}^{dn}$, it holds that $f_p^i(x)\in\ri\left(\mathcal{T}(x_i,\mathcal{H}(\mathcal{C}_p^i(x)))\right)$.
\end{assum}

We next present a uniformly asymptotic agreement result based on the relative interior condition of a tangent cone formed by the supporting hyperrectangle.
\begin{pro}\label{pro}
Suppose $\mathcal{S}_0$ is compact and that Assumptions \ref{assm0}, \ref{assm1}, and \ref{assm-hyper} hold. Then, the cooperative multi-agent system \eqref{eq:switch} achieves uniformly asymptotic agreement on $\mathcal{S}_0$ if and only if its interaction graph $\mathcal{G}_{\sigma(t)}$ is uniformly jointly quasi-strongly connected.
\end{pro}

The proofs of Theorem \ref{thm1} and Proposition \ref{pro} is deferred to Section \ref{sec:cooperative}.

\begin{figure}
\begin{center} \includegraphics[scale=0.35]{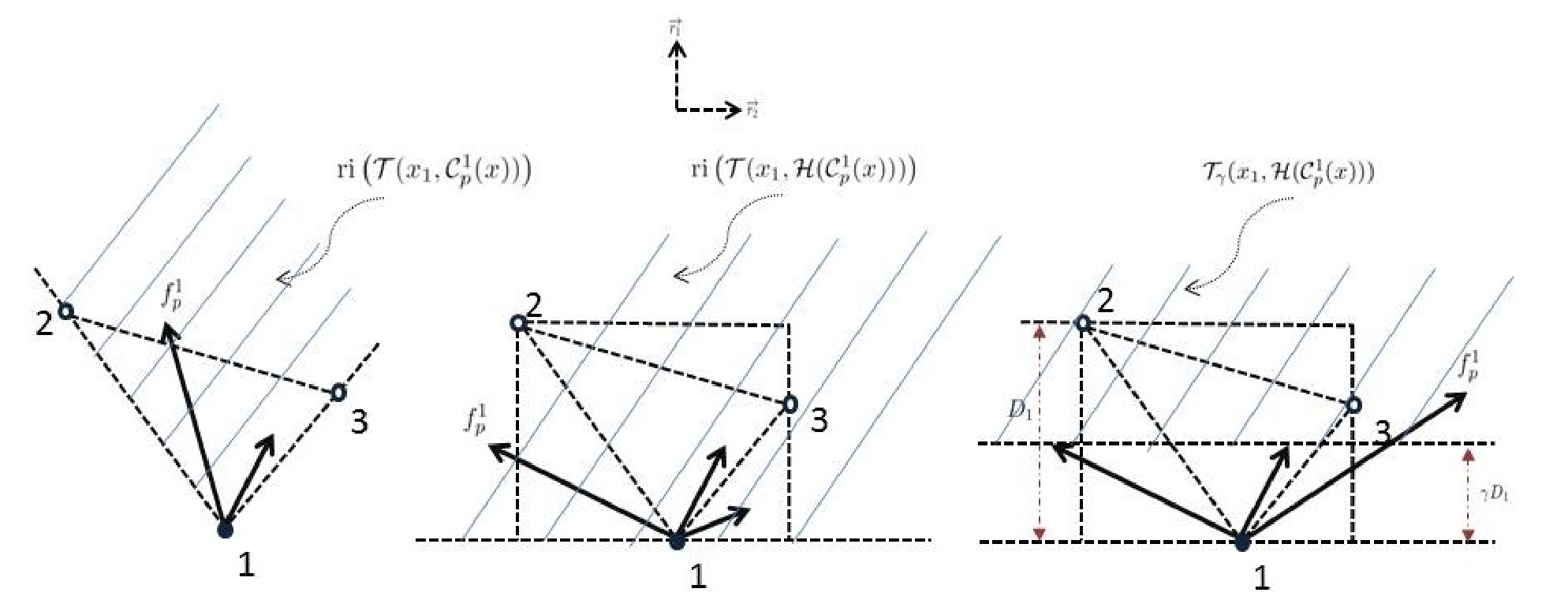}
\caption{The relative interiors of a tangent cone of the convex hull and the supporting hyperrectangle, and $\gamma-$strict tangent cone of the supporting hyperrectangle.}\label{fig-strict}
 \end{center}
\end{figure}

Figure \ref{fig-strict} illustrates the relative interior of a tangent cone of the convex hull (Assumption A2 of \cite{LinZhiyun_SIAM07}), relative interior of a tangent cone of the supporting hyperrectangle (Assumption \ref{assm-hyper}), and strict tangent cone of the supporting hyperrectangle (Assumption \ref{assm-strict}). It is obvious that the vector fields can be chosen  more freely under Assumption \ref{assm-hyper} than  under Assumption A2 of \cite{LinZhiyun_SIAM07}. On the other hand, strict tangent cone condition is a more strict condition than the relative interior condition of a tangent cone. However, exponential agreement can be achieved under strict tangent cone condition while only uniformly asymptotic agreement is achieved under the relative interior condition of a tangent cone.

\begin{remark}
Theorem \ref{thm1} and Proposition \ref{pro} are consistent with the main results in \cite{LinZhiyun_SIAM07,Baras_TAC13,Moreau_TAC05}. Our analysis relies on some critical techniques developed in \cite{LinZhiyun_PhD,LinZhiyun_SIAM07}. Proposition \ref{pro} allows that the vector field belongs to a larger set compared with the convex hull condition proposed in \cite{LinZhiyun_SIAM07,Baras_TAC13,Moreau_TAC05}. In addition, we allow the agent dynamics to switch over a possibly infinite set and we show exponential agreement and derive in the proof of Theorem \ref{thm1} the explicit exponential convergence rate. It follows that by sharing reference directions in addition to the available local information, agreement of multi-agent systems has an enlarged set of interactions and faster convergence speed compared with the case of using only local information.
\end{remark}

To further illustrate Assumptions  \ref{assm-strict} and \ref{assm-hyper}, we discuss two examples.

\begin{figure}
\begin{center}
\includegraphics[scale=0.3]{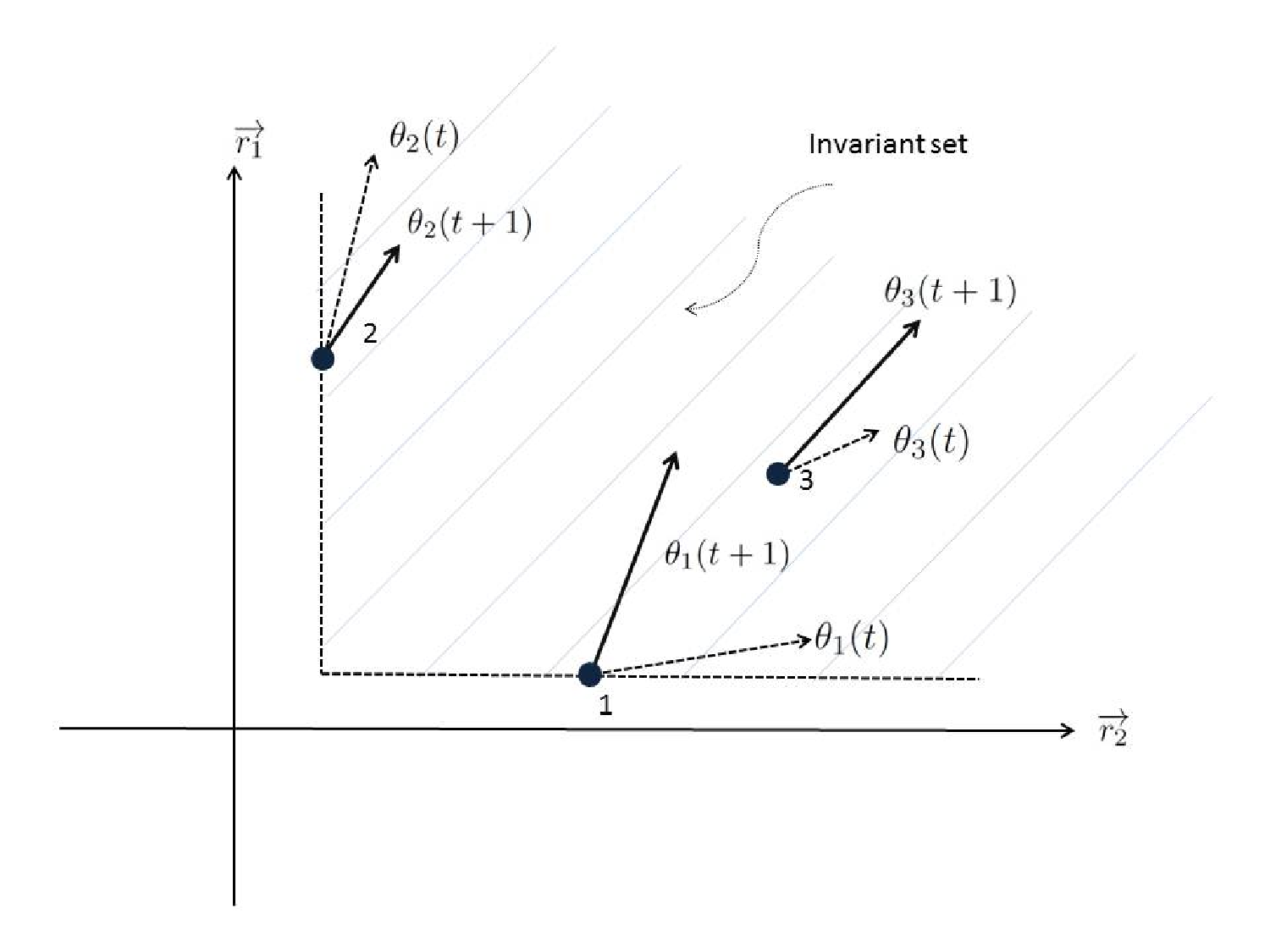}
\caption{An invariant set of Vicsek's model.}\label{fig-Vicsek}
 \end{center}
\end{figure}

\noindent{\bf Example 1.} Let us first consider Vicsek's model \cite{VicsekEtAl95}.
In particular, consider agent $i\in\mathcal{V},$ moving in the plane with position $(x_i(t),y_i(t))$, the same absolute velocity $v$, and the heading $\theta_i(t)$ at discrete time $t=0,1,\dots$. The position and angle updates are described by
\begin{align}
x_i(t+1)&=x_i(t)+v\cos\theta_i(t),\label{eq:xi}\\
y_i(t+1)&=y_i(t)+v\sin\theta_i(t),\label{eq:yi}\\
\theta_i(t+1)&=\arctan\frac{\sum_{j\in\mathcal{N}_i(t)}\sin\theta_j(t)}{\sum_{j\in\mathcal{N}_i(t)}\cos\theta_j(t)},\label{eq:theta}
\end{align}
for all $i\in \mathcal{V}$,
where by convention it is assumed that $i\in \mathcal{N}_i(t)$. From \eqref{eq:theta}, we see that Vicsek's model inherently uses a ``compass''-like directional information. Then, similar to the analysis of Theorem \ref{thm1}, we can easily show that the first quadrant is an invariant set for \eqref{eq:xi} and \eqref{eq:yi}. This can be verified by the fact that $\theta_i(t+1)\in[0,\frac{\pi}{2}]$ when $\theta_j(t)\in [0,\frac{\pi}{2}]$ for all $j\in\mathcal{N}_i(t)$. Figure \ref{fig-Vicsek} illustrates this point for three agents. At time $t$, the vector filed of all the agents are pointing inside the first quadrant, so  agents construct an ``unbounded'' hyperrectangle (both the upper and right bounds are at infinity).
This ``unbounded'' hyperrectangle is the invariant set for the positions of all the agents. The existence of left and lower bounds of the hyperrectangle guarantees that agents 1 and 2 satisfy Assumption \ref{assm-strict}.
However, it is easy to verify that agent~3 does not satisfy Assumption \ref{assm-strict} since the upper and right bounds of the hyperrectangle do not exist. Therefore, position agreement cannot be achieved in general for Vicsek's model.

\noindent{\bf Example 2.} Consider the following dynamics for each agent $i\in \mathcal{V}$:
\begin{equation}
\dot x_i=f_{\sigma(t)}^i(x)=R_{\sigma(t)}^i(x)\sum_{j\in \mathcal{N}_i(\sigma(t))} a_{ij}(x) \big(x_j-x_i\big),\label{eq:rotation-type}
\end{equation}
where $a_{ij}(x)>0$ is a continuous function representing the weight of arc $(j,i)$, and  $R_{\sigma(t)}^i(x)\in \mathbb{R}^{d\times d}$ is a state-dependent rotation matrix which is continuous in $x$ for any fixed $\sigma\in\mathfrak{P}$. Certainly the dynamics described in (\ref{eq:rotation-type}) is beyond the convex hull agreement protocols
\cite{LinZhiyun_SIAM07,Baras_TAC13,Moreau_TAC05}. With the results in Theorem \ref{thm1} and Proposition \ref{pro}, it becomes evident that the existence of  $R_{\sigma(t)}^i(x)$ may still guarantee agreement as long as $R_{\sigma(t)}^i(x)$ rotates the convex hull vector filed, $\sum_{j\in \mathcal{N}_i(\sigma(t))} a_{ij}(x) \big(x_j-x_i\big)$, within the proposed tangent cones given by the local supporting hyperrectangle. Certainly this does not mean that $R_{\sigma(t)}^i(x)$ should be sufficiently small since from Figure \ref{fig-strict} this rotation angle can be large for proper $x$ under certain interaction rules. This can also be viewed as a structural robustness of the proposed ``compass''-based framework.

\subsection{Cooperative-antagonistic networks}
Next, we study the convergence property of the  cooperative--antagonistic networks. Define $\overline{\mathcal{C}}_p^i(x):=\co\{x_i,x_j\sgn^{ij}_p:j\in \mathcal{N}_i(p)\}$. We impose the following assumption.
\begin{assum}\label{assm3}
For all $i\in \mathcal{V}$, $p\in \mathfrak{P}$ and $x\in\mathbb{R}^{dn}$, it holds that $f_p^i(x)\in \mathcal{T}_\gamma(x_i,\mathcal{H}(\overline{\mathcal{C}}_p^i(x)))$.
\end{assum}

Assumption \ref{assm3} follows the model for antagonistic interactions introduced in \cite{Altafini_TAC13}, where simple examples can be found on that state agreement cannot always be achieved for cooperative--antagonistic networks. Instead, it is possible that agents converge to
values with opposite signs, which is known as bipartite consensus \cite{Altafini_TAC13}.
We present the following result for cooperative--antagonistic networks.

\begin{them}\label{thm2}
Let Assumptions \ref{assm0}, \ref{assm1} and \ref{assm3} hold. Then, if (and in general only if) the interaction graph $\mathcal{G}_{\sigma(t)}$ is uniformly jointly strongly connected, all the agents' trajectories asymptotically converge  for cooperative-antagonistic multi-agent system \eqref{eq:switch}, and their limits agree componentwise in absolute values for every initial time and initial state.
\end{them}

Here by ``in general only if,'' we mean that we can always construct simple examples with fixed interaction rule, for which strong connectivity is necessary for the result in Theorem \ref{thm2} to stand. The proof of Theorem \ref{thm2} will be presented in Section \ref{sec:antagonistic}.
Compared with the results given in \cite{Altafini_TAC13}, Theorem \ref{thm2} requires no conditions on the structural balance of the network. Theorem \ref{thm2} shows that every positive or negative arc contributes to the convergence of the absolute values of the nodes' states, even for general nonlinear multi-agent dynamics.

The exponential agreement and uniformly asymptotical agreement results given in Theorem~\ref{thm1} and Proposition \ref{pro} rely on uniformly jointly quasi-strong connectivity, while the result in Theorem~\ref{thm2} needs uniformly jointly strong connectivity.
For cooperative networks,
we establish the exponential convergence rate in the proof of Theorem \ref{thm1}. In contrast, for cooperative--antagonistic networks in Theorem \ref{thm2}, the convergence speed is unclear. We conjecture that exponential convergence might not hold in general under the conditions of Theorem \ref{thm2}. The reason is that Lemmas~\ref{lem:bound2} and \ref{lem:contraction2} given in Section \ref{sec:cooperative} cannot be recovered for cooperative--antagonistic networks.

We believe that differences between Theorems \ref{thm1} and \ref{thm2} discussed in the previous remarks reveal some important distinctions of cooperative and cooperative--antagonistic networks.

\section{Cooperative Multi-agent Systems}\label{sec:cooperative}

In this section, we focus on the case of cooperative multi-agent systems. We will prove Theorem~\ref{pro} and Proposition~\ref{thm1} by analyzing a contraction property of \eqref{eq:switch}, with the help of a series of preliminary lemmas.

\subsection{Invariant set}\label{sec:invariant1}

We introduce the following definition.
\begin{defn}\label{def:invariant}
A set $\mathcal{M}\subset \mathbb{R}^{dn}$ is an invariant set for the
system \eqref{eq:switch} if for all $t_0\geq 0$,
$
x(t_0)\in \mathcal{M}\quad \Longrightarrow \quad x(t)\in \mathcal{M},~\forall t\geq t_0.
$
\end{defn}

For all $k\in\mathcal{D}$, define
$
M_k(x(t))=\max_{i\in\mathcal{V}}\{x_{ik}(t)\}, ~~m_k(x(t))=\min_{i\in\mathcal{V}}\{x_{ik}(t)\},
$
where $x_{ik}$ denotes $k\th$ entry of $x_i$. In addition, define the supporting hyperrectangle by the initial states of all agents as $\mathcal{H}_0:=\mathcal{H}(\mathcal{C}(x(t_0)))$, where $\mathcal{C}(x)=\co\{x_1,x_2,\dots,x_n\}$.

In the following lemma, we show that the supporting hyperrectangle formed by the initial states of all agents is an invariant set for system \eqref{eq:switch}.
\begin{lem}\label{lem:invariant1}
Let Assumptions \ref{assm0}, \ref{assm1}, and \ref{assm-strict} or Assumptions \ref{assm0}, \ref{assm1}, and \ref{assm-hyper} hold. Then, $\mathcal{H}_0^n$ is an invariant set,  i.e., $x_i(t)\in \mathcal{H}_0$, $\forall i\in \mathcal{V}$, $\forall t\geq t_0$.
\end{lem}

{\it Proof.}
We first show that $D^+M_k(t)\leq 0$, $\forall k\in\mathcal{D}$.
Let $\widehat{\mathcal{V}}(t)=\{i\in\mathcal{V}: x_{ik}(t)=M_k(t)\}$ be the set of indices where the maximum is reached at $t$.
It then follows from Lemma \ref{lem:Dini} that for all $k\in \mathcal{D}$,
$
D^+M_k(t)=\max_{i\in\widehat{\mathcal{V}}(t)}\dot x_{ik}= \max_{i\in\widehat{\mathcal{V}}(t)} f^{ik}_{\sigma(t)}(x(t)),
$
where $f^{ik}_{\sigma(t)}$ denotes $k\th$ entry of the vector $f^{i}_{\sigma(t)}$.
Consider any initial state $x(t_0)\in \mathcal{H}_0^n$ and any initial time $t_0$.
It follows from Definition \ref{def:cone} and Lemma \ref{lem:cone} that
$
f_p^i(x)\in\mathcal{T}_\gamma(x_i,\mathcal{H}(\mathcal{C}_p^i(x)))\subseteq\mathcal{T}(x_i,\mathcal{H}(\mathcal{C}_p^i(x))),
~ \forall i\in \mathcal{V}, ~\forall p\in \mathfrak{P},
$
for Assumption \ref{assm-strict} and
$
f_p^i(x)\in\ri\left(\mathcal{T}(x_i,\mathcal{H}(\mathcal{C}_p^i(x)))\right)\subseteq\mathcal{T}(x_i,\mathcal{H}(\mathcal{C}_p^i(x))),
~ \forall i\in \mathcal{V}, ~\forall p\in \mathfrak{P},
$
for Assumption \ref{assm-hyper}.
It follows from the definition of the tangent cone that $f_p^{ik}(x)\leq 0$ for all $i\in \mathcal{V}$ satisfying $x_{ik}=M_k$. It follows that for all $k\in \mathcal{D}$ and any $x\in \mathcal{H}_0^n$,
$
D^+M_k(t)\leq 0.
$
We can similarly show that for all $k\in \mathcal{D}$, $D^+m_k(t)\geq 0$.

Therefore, it follows that $m_k(x(t_0))\leq x_{ik}(t)\leq M_k(x(t_0))$, $\forall k\in\mathcal{D}$, $\forall i\in \mathcal{V}$, $\forall t\geq t_0$.
Then, based on the definition of $\mathcal{H}_0$, we have shown that $\mathcal{H}_0$ is an invariant set.
\hfill$\square$

\subsection{Interior agents}\label{sec:interior1}
\begin{figure}
\begin{center}
\includegraphics[scale=0.26]{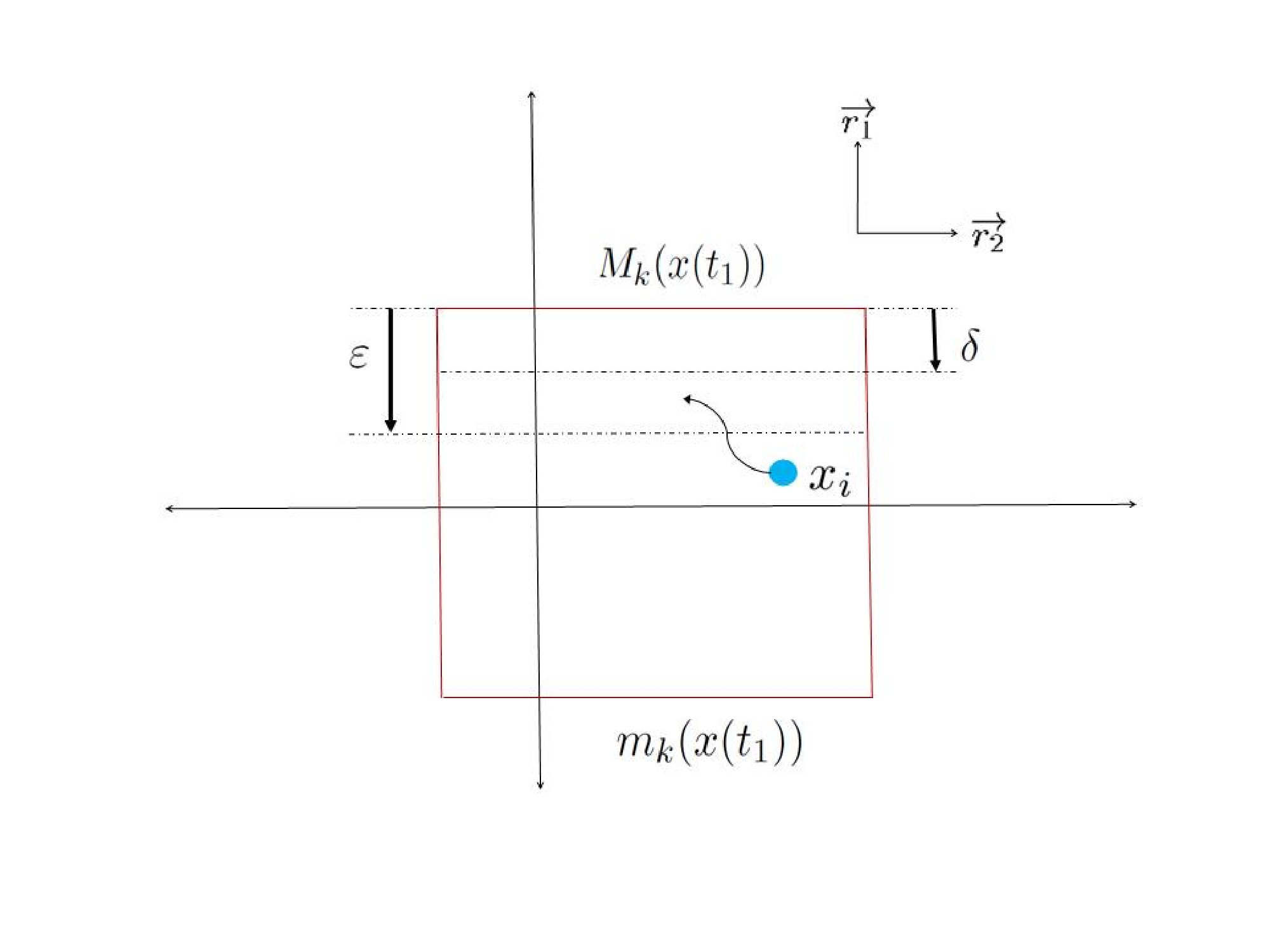}
\caption{Illustration of Lemma \ref{lem:bound1}}\label{fig-interior}
 \end{center}
\end{figure}

In this subsection, we study the state evolution of the agents whose states are interior points of $\mathcal{H}(\mathcal{C}(x))$.
In the following lemma, we show that the projection of the state on any coordinate axis is strictly less than an explicit upper bound as long as it is initially strictly less than this upper bound. Figure \ref{fig-interior} illustrates the following Lemma \ref{lem:bound1}.

The proof follows from a similar argument used in the proof Lemma 4.9 in \cite{LinZhiyun_PhD} and the following lemma holds separately for any $k\in\mathcal{D}$.
\begin{lem}\label{lem:bound1}
Let Assumptions \ref{assm0}, \ref{assm1}, and \ref{assm-strict} or Assumptions \ref{assm0}, \ref{assm1}, and \ref{assm-hyper} hold. Also assume that $\mathcal{G}_{\sigma(t)}$ is uniformly jointly quasi-strongly connected. Fix any $k\in\mathcal{D}$. For any $(t_1,x(t_1))\in \mathbb{R}\times \mathcal{H}_0^n$, any $\varepsilon>0$, and any $T^*>0$, if $x_{ik}(t_2)\leq  M_k(x(t_1))-\varepsilon$ at some $t_2\geq t_1$, then
$x_{ik}(t)\leq M_k(x(t_1))-\delta$, where $\delta=e^{-L_1^* T^*}\varepsilon$ for all $t\in[t_2,t_2+T^*]$, and $L_1^*$ is a positive constant related to $\mathcal{H}_0$.
\end{lem}

{\it Proof.}
Fix $(t_1,x(t_1))\in \mathbb{R}\times \mathcal{H}_0^n$ and any $k\in\mathcal{D}$. Denote $\psi=x(t_1)$ and
$\mathcal{M}_{ik}=\mathcal{H}(\psi)\times\dots\mathcal{H}(\psi)\times\underbrace{\mathcal{H}^o_k(\psi)}_{{\rm the} ~i {\rm th~ entry}}
\times\mathcal{H}(\psi)\times\dots\times \mathcal{H}(\psi),
$
where $\mathcal{H}(\psi)=[m_1(\psi),M_1(\psi)]\times\dots \times[m_d(\psi),M_d(\psi)]$, and $
\mathcal{H}^o_k(\psi)=[m_1(\psi),M_1(\psi)]\times\dots   \times [m_{k-1}(\psi),M_{k-1}(\psi)]
\times[m_{k+1}(\psi),M_{k+1}(\psi)]\times\dots \times[m_d(\psi),M_d(\psi)].$
The rest of the proof will be divided in three steps.

\noindent{ \it (Step I).} Define the following nonlinear function
\begin{align}
g_{\psi,k}(\chi):[m_k(\psi), M_k(\psi)] \rightarrow \mathbb{R},~~
\chi \mapsto
\sup_{p\in\mathfrak{P}}\{\max_{i\in\mathcal{V}}\{\max_{y\in \mathcal{M}_{ik}}\{f_p^{ik}(x_{ik},y):x_{ik}=\chi\}\}\},
\end{align}
where $f_p^{ik}(x_{ik},y)$ denotes the $k\th$ entry of the vector $f_p^{i}(x)$, $x_{ik}$ denotes the $k\th$ entry of the vector $x_i$ and $y$ denotes all the other components of $x$ except $x_{ik}$. The nonlinear function $g_{\psi,k}(\chi)$ is used as an upper bound of $f_{\sigma(t)}^{ik}(x)$ and the argument $\chi$ is used to describe the state $x_{ik}$.
In this step, we establish some useful properties of $g_{\psi,k}(\cdot)$  based on Lemmas \ref{lem:Lipschitz1-1} and \ref{lem:Lipschitz1-2} in the Appendices. We make the following claim.

Claim A: (i) $g_{\psi,k}(\chi)=0$ if $\chi=M_k(\psi)$; (ii) $g_{\psi,k}(\chi)>0$ if $\chi\in [m_k(\psi), M_k(\psi))$; (iii) $g_{\psi,k}(\chi)$ is Lipschitz continuous with respect to $\chi$ on $[m_k(\psi), M_k(\psi)]$.

It follows from Definition \ref{def:cone}, Lemma \ref{lem:cone} and the similar analysis of Lemma \ref{lem:invariant1} (by replacing $t_0$ with $t_1$) that $\forall t\geq t_1$,
$f_p^i(x)\in \mathcal{T}_\gamma(x_i,\mathcal{H}(\mathcal{C}_p^i(x))) ({\rm or~}\ri\left(\mathcal{T}(x_i,\mathcal{H}(\mathcal{C}_p^i(x)))\right) )
\subseteq
\mathcal{T}(x_i,\mathcal{H}(\mathcal{C}_p^i(x))) \subseteq \mathcal{T}(x_i,\mathcal{H}(\mathcal{C}(x)))
\subseteq  \mathcal{T}(x_i,\mathcal{H}(\mathcal{C}(\psi))),
~\forall i\in \mathcal{V}, ~\forall p\in \mathfrak{P}$. Then, it follows from Definition \ref{def:cone} that $f_p^{ik}(x)\leq 0$ when $x_{ik}=M_k(\psi)$. This implies that $g_{\psi,k}(\chi)\leq 0$ when $\chi=M_k(\psi)$ based on the definition of $g_{\psi,k}(\chi)$. We next show that actually $g_{\psi,k}(\chi)=0$ when $\chi=M_k(\psi)$. Since $\mathcal{G}_{\sigma(t)}$ is uniformly jointly quasi-strongly connected, there must exist a $\bar{p}\in\mathfrak{P}$ such that $\mathcal{G}_{\bar {p}}$ has a nonempty arc set $\mathcal{E}_{\bar{p}}$. We can then choose $\bar{i}\in\mathcal{V}$ and $\bar{p}$ such that agent $\bar{i}$ has at least one neighbor agent, i.e., $\mathcal{N}_{\bar{i}}(\bar{p})$ is not empty since $\mathcal{E}_{\bar{p}}$ is nonempty. We next choose $x_j=x_{\bar{i}}\in \mathcal{H}(\mathcal{C}(\psi))$, for all $j\in \mathcal{N}_{\bar{i}}(\bar{p})$, where $x_{\bar{i}k}=M_k(\psi)$. In such a case, $\mathcal{H}(\mathcal{C}_{\bar{p}}^{\bar{i}}(x))$ is the singleton $\{x_{\bar{i}}\}$ and it follows from Assumption \ref{assm-strict} (or \ref{assm-hyper}) that $f_{\bar{p}}^{\bar{i}}(x)=0$. Therefore, based on the definition of $g_{\psi,k}(\chi)$, we know that $g_{\psi,k}(\chi)=0$ if $\chi=M_k(\psi)$. This proves (i).

Next, for any $\chi\in [m_k(\psi), M_k(\psi))$, we still use the same $\bar{p}$ and $\bar{i}$ as those in the proof of Claim A(i). We choose $x_{\bar{i}k^o}=M_{k^o}(\psi)$, $\forall k^o\in\{1,\dots,k-1,k+1,\dots,d\}$ and  $x_{jk}=M_k(\psi)$, $\forall k\in\mathcal{D}$, for all $j\in \mathcal{N}_{\bar{i}}(\bar{p})$. Note that $x_{\bar{i}k}=\chi<M_k(\psi)$.
In such a case, $\mathcal{H}(\mathcal{C}_{\bar{p}}^{\bar{i}}(x))$ is a line from point $
(M_1(\psi),\dots,M_{k-1}(\psi),\chi,M_{k+1}(\psi),\dots,M_d(\psi))
$ to $(M_1(\psi),M_2(\psi),\dots,M_d(\psi))$.
It then follows from Assumption \ref{assm-strict} that $f_{\bar{p}}^{\bar{i}}(x)\geq \gamma(M_{k}(\psi)-\chi)>0$ or from Assumption \ref{assm-hyper} that $f_{\bar{p}}^{\bar{i}}(x)>0$. This verifies that $g_{\psi,k}(\chi)>0$, $\forall \chi\in [m_k(\psi), M_k(\psi))$. This proves (ii).

Finally, it follows from Lemma \ref{lem:Lipschitz1-2} that $g^{ik}_p(x_{ik}):[m_k(\psi), M_k(\psi)] \rightarrow \mathbb{R}$, $x_{ik}\mapsto\max_{y\in \mathcal{M}_{ik}}f_p^{ik}(x_{ik},y)$ is locally Lipschitz with respect to $x_{ik}$, $\forall k\in\mathcal{D}$, $\forall i\in\mathcal{V}$ and $\forall p\in\mathfrak{P}$. Then, it follows from Theorem~1.14 of \cite{Markley_Book} that $g^{ik}_p(x_{ik})$ is (globally) Lipschitz continuous with respect to $x_{ik}$ on $[m_k(\psi), M_k(\psi)]$. From the first property of $g_{\psi,k}(\chi)$, it follows that $g_{\psi,k}(M_k(\psi))=0$. Therefore, based on Lemma \ref{lem:Lipschitz1-1},
it follows that $g_{\psi,k}(\chi)$ is Lipschitz continuous with respect to $\chi$ on $[m_k(\psi), M_k(\psi)]$. This proves (iii) and the claim holds.

\noindent {\it (Step II).}
In this step, we construct and investigate the nonlinear function $h_{\mathcal{H}_0,k}(\cdot)$, which is derived by  $g_{\psi,k}(\cdot)$ with the argument $\varphi=\chi-M_k(\psi)$ measuring the difference between $x_{ik}$ and the upper boundary $M_k(\psi)$. Define
\begin{align}
h_{\mathcal{H}_0,k}(\varphi):[\hat{a}_k-\breve{a}_k,0]\rightarrow \mathbb{R}, ~~
\varphi\mapsto \begin{cases}
g_{\psi,k}(\varphi+M_k(\psi)); \quad \text{if}~~ \varphi\in[m_k(\psi)-M_k(\psi), 0]\\
g_{\psi,k}(m_k(\psi)); \quad \text{if}~~ \varphi\in[\hat{a}_k-\breve{a}_k,m_k(\psi)-M_k(\psi)),
\end{cases}
\end{align}
where $\hat{a}_k=m_k(x(t_0))$ and $\breve{a}_k=M_k(x(t_0))$ are constants determined by $\mathcal{H}_0$. Obviously, $h_{\mathcal{H}_0,k}(\varphi)$ is continuous. We make the following claim.

Claim B:  (i) $h_{\mathcal{H}_0,k}(\varphi)$ is Lipschitz continuous with respect to $\varphi$ on $[\hat{a}_k-\breve{a}_k,0]$, where the Lipschitz constant is denoted by $L_1^*$ and $L_1^*$ is related to the initial bounded set $\mathcal{H}_0$; (ii) $h_{\mathcal{H}_0,k}(\varphi)>0$ if $\varphi\in [\hat{a}_k-\breve{a}_k,0)$; (iii) $h_{\mathcal{H}_0,k}(\varphi)=0$ if $\varphi=0$.

Note that $h_{\mathcal{H}_0,k}(\varphi)=g_{\psi,k}(\varphi+M_k(\psi))$ is compact on the compact set $[m_k(\psi)-M_k(\psi),0]$. It follows that $h_{\mathcal{H}_0,k}(\varphi)$ is Lipschitz continuous with respect to $\varphi$ on the compact set $[\hat{a}_k-\breve{a}_k,0]$.  This shows that (i) holds and properties (ii) and (iii) follow directly from the definition of $h_{\mathcal{H}_0,k}(\varphi)$.

\noindent {\it (Step III).} In this step, we take advantage of $g_{\psi,k}(\chi)$ and $h_{\mathcal{H}_0,k}(\varphi)$ to show that $x_{ik}$ will be always strictly less than the upper bound $M_k(\psi)$ as long as it is initially strictly less than $M_k(\psi)$.

Suppose $x_{ik}(t_2)\leq  M_k(\psi)-\varepsilon$ at some $t_2\geq t_1$ and let $T^*>0$. Based on the definition of $g_{\psi,k}(\chi)$, it follows that
$
\dot x_{ik}(t)=f_{\sigma(t)}^{ik}(x(t))\leq g_{\psi,k}(x_{ik}(t)),~\forall t\geq t_2.
$
Let $\chi(t)$ be the solution of $\dot\chi=g_{\psi,k}(\chi)$ with initial condition $\chi(t_2)=x_{ik}(t_2)$. Based on the Comparison Lemma (Lemma 3.4 of \cite{Khalil_book}), it follows that $x_{ik}(t)\leq \chi(t)$, $\forall t\geq t_2$.

Note that $\varphi=\chi-M_k(\psi)$ and $\dot\varphi=g_{\psi,k}(\chi)=h_{\mathcal{H}_0,k}(\varphi)$. It follows from the first property of $h_{\mathcal{H}_0,k}(\varphi)$ that $|h_{\mathcal{H}_0,k}(\varphi)-h_{\mathcal{H}_0,k}(0)|\leq L_1^*|\varphi|$, $\forall \varphi\in [\hat{a}_k-\breve{a}_k,0]$. This shows that $h_{\mathcal{H}_0,k}(\varphi)\leq -L_1^*\varphi$ based on the second and the third properties of $h_{\mathcal{H}_0,k}(\varphi)$. Thus, the solution of $\dot\varphi=h_{\mathcal{H}_0,k}(\varphi)$ satisfies $\varphi(t)\leq e^{-L_1^*(t-t_2)}\varphi(t_2)$, $\forall t\geq t_2$ based on the Comparison Lemma.

Therefore, $x_{ik}(t)\leq \chi(t)=\varphi(t)+M_k(\psi)\leq e^{-L_1^*(t-t_2)}(\chi(t_2)-M_k(\psi))+M_k(\psi)\leq e^{-L_1^*T^*}(x_{ik}(t_2)-M_k(\psi))+M_k(\psi)\leq M_k(\psi)-e^{-L_1^*T^*}\varepsilon$ for all $t\in[t_2,t_2+T^*]$.
\hfill$\square$

The following lemma is symmetric to Lemma \ref{lem:bound1}. The proof can be obtained using the proof of Lemma \ref{lem:bound1} under the transformation $z_i=-x_i,i=1,\dots,n$ and it is therefore omitted.
\begin{lem}\label{lem:bound2}
Let Assumptions \ref{assm0}, \ref{assm1}, and \ref{assm-strict} or Assumptions \ref{assm0}, \ref{assm1}, and \ref{assm-hyper} hold. Also assume that $\mathcal{G}_{\sigma(t)}$ is uniformly jointly quasi-strongly connected. Fix any $k\in\mathcal{D}$. For any $(t_1,x(t_1))\in \mathbb{R}\times \mathcal{H}_0^n$, any $\varepsilon>0$, and any $T^*>0$, if $x_{ik}(t_2)\geq m_k(x(t_1))+\varepsilon$ at some $t_2\geq t_1$, then
$x_{ik}(t_2)\geq m_k(x(t_1))+\delta$, where $\delta=e^{-L_2^*T^*}\varepsilon$ for all $t\in[t_2,t_2+T^*]$, where $L_2^*$ is a positive constant related to $\mathcal{H}_0$.
\end{lem}

\subsection{``Boundary'' agents}\label{sec:boundary1}

\begin{figure}
\begin{center}
\includegraphics[scale=0.26]{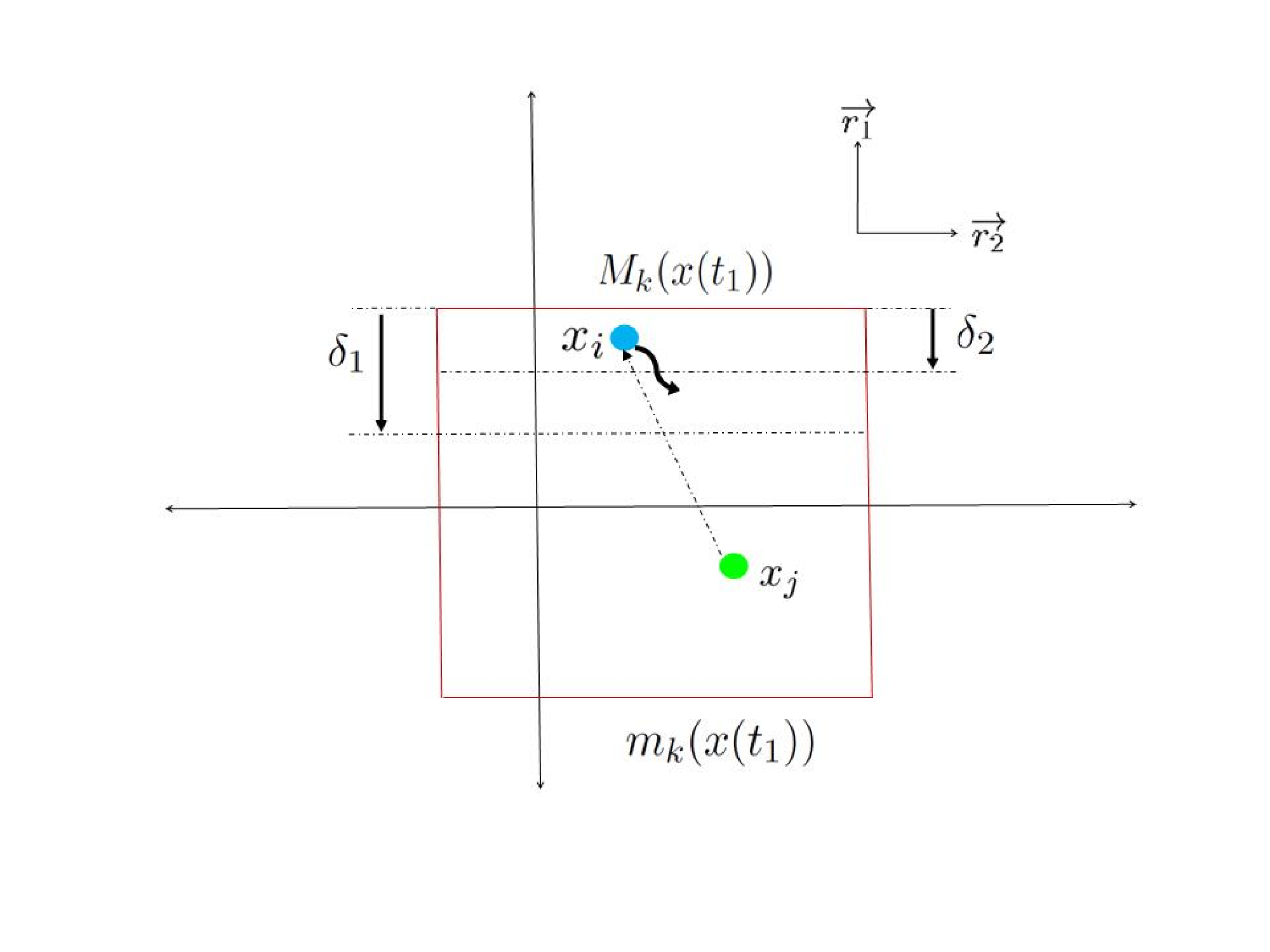}
\caption{Illustration of Lemma \ref{lem:contraction1}}\label{fig-boundary}
 \end{center}
\end{figure}

In the following lemma, we show that any agent that is attracted by an ``interior'' agent will become an ``interior'' agent after a finite time period. Figure~\ref{fig-boundary} illustrates Lemma~\ref{lem:contraction1}.

\begin{lem}\label{lem:contraction1}
Let Assumptions \ref{assm0} and \ref{assm1} hold and assume that $\mathcal{G}_{\sigma(t)}$ is uniformly jointly quasi-strongly connected. Fix any $k\in\mathcal{D}$. For any $(t_1,x(t_1))\in \mathbb{R}\times \mathcal{H}_0^n$, any $\delta_1>0$ and any $T^*>0$, assume that there is an arc $(j, i)$ and a time $t_2\geq t_1$ such that $ j\in \mathcal{N}_i(\sigma(t))$, and $x_{jk}(t)\leq M_k(x(t_1))-\delta_1$ for all $t\in[t_2,t_2+\tau_d]$. Then, there exists a $t_3\in[t_1,t_2+\tau_d]$ such that $x_{ik}(t)\leq M_k(x(t_1))-\delta_2$, for all $t\in[t_3,t_3+T^*]$. Here, if  Assumption \ref{assm-strict} is satisfied, $\delta_2=e^{-L_1^*T^*}\min\{\frac{\gamma\tau_d\delta_1}{L_1^+\tau_d+1},\delta_1\}$ for some positive constants $L_1^*$ and $L_1^+$ related to $\mathcal{H}_0$. If Assumption \ref{assm-hyper} is satisfied, $\delta_2=e^{-L_1^*T^*}\upsilon(\delta_1)$ for some positive constant $L_1^*$ and a continuous positive-definite function $\upsilon(\cdot)$ both related to $\mathcal{H}_0$.
\end{lem}

{\it Proof.}
We first show that there exists $t_3\in[t_1,t_2+\tau_d]$ such that $x_{ik}(t_3)\leq M_k(x(t_1))-\varepsilon$, where $\varepsilon=\min\{\frac{\gamma\tau_d\delta_1}{L_1^+\tau_d+1},\delta_1\}$ given Assumption \ref{assm-strict} satisfied or  $\varepsilon=\upsilon_{\mathcal{H}_0}(\delta_1)$ given Assumption \ref{assm-hyper} satisfied. This is
equivalent to show that $\|x_i(t_3)\|_{\mathcal{B}}=0$, where $\mathcal{B}:=\mathcal{H}_{\varepsilon}^k(\mathcal{C}(x(t_1)))$ and an axis-aligned hyperrectangle $\mathcal{H}_{\varepsilon}^k$ defined as $\mathcal{H}_{\varepsilon}^k(\mathcal{C}(x))=\{y\in\mathcal{H}(\mathcal{C}(x)):y_k\leq M_k(x)-\varepsilon\}$. Obviously, $\mathcal{B}$ is compact convex set. Suppose $\|x_i(t_3)\|_{\mathcal{B}}\neq 0$. It then follows that $0<\|x_i(t)\|_\mathcal{B}\leq \varepsilon$ for all $t\in[t_1,t_2+\tau_d]$.

Considering the time interval $t\in[t_2,t_2+\tau_d]$, we define $\overline{x}(t)=[\overline{x}_{11},\dots,\overline{x}_{1d},\overline{x}_{21},
\dots,\overline{x}_{2d},\dots,\\
 \overline{x}_{n1},\dots,\overline{x}_{nd}],$ $\overline{x}_{ik}(t)=M_k(x(t_1))$ for given $i$ and $k$, and $\overline{x}_{i^ok^o}(t)=x_{i^ok^o}(t)$ for $i^o\in \mathcal{V}\setminus \{i\}$ and $k^o\in \mathcal{D}\setminus \{k\}$.
The rest of the proof will be divided into three steps.

\noindent {\em (Step I).}
It has been shown that $f_p^{ik}(x(t))$ is uniformly locally
Lipschitz with respect to $x$ and compact on $\mathcal{H}_0^n$, $\forall i\in \mathcal{V}$, $\forall p\in\mathfrak{P}$ based on Assumption \ref{assm1} and Lemma \ref{lem:invariant1}.
Therefore,
there exists a positive constant $L_1^+$ related to $\mathcal{H}_0$ such that $|f^{ik}_p(\overline{x})|-|f^{ik}_p(x)|\leq |f^{ik}_p(\overline{x})-f^{ik}_p(x)|\leq L_1^+\|x(t)-\overline{x}(t)\|\leq L_1^+\varepsilon$, $\forall p\in\mathfrak{P}$, and $\forall x,\overline{x}\in\mathcal{H}_0^n$.

\noindent {\em (Step II - Assumption \ref{assm-strict}).} In this step, we show that the derivative of $\|x_i(t)\|_\mathcal{B}$ along the solution of \eqref{eq:switch} has a lower bound. For any $p^*\in \mathfrak{P}$ such that there is an arc $( j, i)$ where $ j\in \mathcal{N}_i(p^*)$, and $x_{jk}\leq M_k(x(t_1))-\delta_1$ during $t\in [t_2,t_2+\tau_d]$, it follows from Assumption \ref{assm-strict} of $f^i_{p^*}(\overline{x})\in \mathcal{T}_\gamma(\overline{x}_i,\mathcal{H}(\mathcal{C}^i_{p^*}(\overline{x})))$ and $\overline{x}_{ik}(t)= M_k(x(t_1))$ that
\begin{align}
|f^{ik}_{p^*}(\overline{x})|
\geq  \gamma D_k (\mathcal{H}(\co\{\overline{x}_i,x_j:j\in \mathcal{N}_i(p^*)\}))
 \geq  \gamma D_k (\mathcal{H}(\co\{\overline{x}_i,x_j\}))
 \geq\gamma\delta_1,
 \end{align}
where the first inequality is based on Assumption \ref{assm-strict} by noting that $\overline{x}_i\in \rb_k\mathcal{H}(\co\{\overline{x}_i,x_j:j\in \mathcal{N}_i(p^*)\})$, and $\rb_k\mathcal{H}(\co\{\overline{x}_i,x_j:j\in \mathcal{N}_i(p^*)\})$ is the facet of $\mathcal{H}(\co\{\overline{x}_i,x_j:j\in \mathcal{N}_i(p^*)\})$ perpendicular
to $\overrightarrow{r_k}$. This together with the preceding deduction $|f^{ik}_p(\overline{x})|-|f^{ik}_p(x)|\leq L_1^+\varepsilon$, $\forall p\in\mathfrak{P}$, and $\forall x,\overline{x}\in\mathcal{H}_0^n$,
implies that $|f^{ik}_{p^*}(x)|
\geq  |f^{ik}_{p^*}(\overline{x})|-L_1^+\varepsilon
\geq~\gamma\delta_1-L_1^+\varepsilon$ for any $p^*\in \mathfrak{P}$ such that there is an arc $( j, i)$ where $ j\in \mathcal{N}_i(p^*)$, and $x_{jk}\leq M_k(x(t_1))-\delta_1$ during $t\in [t_2,t_2+\tau_d]$. Note that  $\varepsilon=\min\{\frac{\gamma\tau_d\delta_1}{L_1^+\tau_d+1},\delta_1\}$ is chosen sufficiently small at the beginning of the proof such that $\gamma\delta_1-L_1^+\varepsilon$ is positive.

Therefore, based on the assumptions of Lemma \ref{lem:contraction1}, it follows that for all $t\in [t_2,t_2+\tau_d]$,
\begin{align}
|D^+\|x_i(t)\|_{\mathcal{\mathcal{B}}}| =&~|\langle \sgn(x_i(t)-P_{\mathcal{B}}(x_i(t))),
f_{\sigma(t)}^{i}(x(t))\rangle|
=|f_{\sigma(t)}^{ik}(x(t))|
\geq~\gamma\delta_1-L_1^+\varepsilon,
 \end{align}
where the componentwise sign function $\sgn(\cdot)$ is defined as $\sgn(z)=[\sgn(z_{1}),\sgn(z_{2}), \dots,\sgn(z_{d})]$ for a vector $z=[z_1,z_2,\dots,z_d]$ and $\sgn(z_1)$ is the sign function: $\sgn(z_1)=1$ if $z_1>0$, $\sgn(z_1)=0$ if  $z_1=0$, and $\sgn(z_1)=-1$ if $z_1<0$.
Note that $\sgn(x_i-P_{\mathcal{B}}(x_i))=\overrightarrow{r_k}$ whenever $\|x_i\|_\mathcal{B}>0$.

\noindent {\em (Step II - Assumption \ref{assm-hyper}).}
Fix $(t_1,x(t_1))\in \mathbb{R}\times \mathcal{H}_0^n$. Denote $\psi=x(t_1)$ and $\mathcal{M}_{ik}=\mathcal{H}(\psi)\times\dots\mathcal{H}(\psi)\times\mathcal{H}^o_k(\psi)
\times\mathcal{H}(\psi)\times\dots\times \mathcal{H}(\psi)$, where $\mathcal{H}(\psi)=[m_1(\psi),M_1(\psi)]\times\dots \times[m_d(\psi),M_d(\psi)]$, and $\mathcal{H}^o_k(\psi)=[m_1(\psi),M_1(\psi)]\times\dots  \times [m_{k-1}(\psi),M_{k-1}(\psi)]
\times[m_{k+1}(\psi),M_{k+1}(\psi)]\times\dots \times[m_d(\psi),M_d(\psi)]$.
Define
\begin{align}
d_{\psi,k}(\delta_1)=\inf_{p\in\mathfrak{P}}\{\min_{i\in\mathcal{V}}\{\min_{y\in \mathcal{U_{\psi}}}\{|f_p^{ik}(M_{k}(\psi),y)|\}\}\},
 \end{align}
where
$\mathcal{U}_{\psi}(i,k,p,\delta_1)=\{y\in\mathcal{M}_{ik}:\exists j\in \mathcal{N}_i(p) {\rm ~such~that~} x_{jk}\leq M_k(\psi)-\delta_1\}$. Based on the relative interior condition of Assumption \ref{assm-hyper}, we know that $d_{\psi,k}(\delta_1)>0$ for $\delta_1>0$.

For any $p^*\in \mathfrak{P}$ such that there is an arc $( j, i)$, where $ j\in \mathcal{N}_i(p^*)$, and $x_{jk}\leq M_k(x(t_1))-\delta_1$, we know from the definition of $d_{\psi,k}(\cdot)$ that for all $t\in [t_2,t_2+\tau_d]$, $|f_{p^*}^{ik}(\overline{x}(t))|\geq d_{\psi,k}(\delta_1)$. This together with the preceding deduction $|f^{ik}_p(\overline{x})|-|f^{ik}_p(x)|\leq L_1^+\varepsilon$, $\forall p\in\mathfrak{P}$, and $\forall x,\overline{x}\in\mathcal{H}_0^n$,
implies that for all $t\in [t_2,t_2+\tau_d]$,
\begin{align}
|D^+\|x_i(t)\|_{\mathcal{\mathcal{B}}}|
=~|f_{\sigma(t)}^{ik}(x(t))|
\geq~d_{\psi,k}(\delta_1)-L_1^+\varepsilon.
 \end{align}

Before moving on, we define $\upsilon_{\mathcal{H}_0,k}(\delta_1):[0,\breve{a}_k-\hat{a}_k]\rightarrow [0,\infty)$,
\begin{align}
\delta_1\mapsto \begin{cases}
\min\left\{\delta_1,\frac{\tau_d d_{\psi,k}(\delta_1)}{\tau_dL_1^++1}\right\}; \quad \text{if}~~ \delta_1\in[0,M_k(\psi)-m_k(\psi)],\\
\min\left\{M_k(\psi)-m_k(\psi),\frac{\tau_d d_{\psi,k}(M_k(\psi)-m_k(\psi))}{\tau_dL_1^++1}\right\};  \quad \text{if}~~ \delta_1\in(M_k(\psi)-m_k(\psi),\breve{a}_k-\hat{a}_k],
\end{cases}
\end{align}
where $\hat{a}_k=m_k(x(t_0))$ and $\breve{a}_k=M_k(x(t_0))$ are constants determined by $\mathcal{H}_0$.
Obviously, $\upsilon_{\mathcal{H}_0,k}(\delta_1)$ is a continuous positive-definite function
since $\upsilon_{\mathcal{H}_0,k}(\delta_1)=0$ for $\delta_1=0$, and $\upsilon_{\mathcal{H}_0,k}(\delta_1)>0$ for $\delta_1>0$. Also note that $\upsilon_{\mathcal{H}_0,k}(\delta_1)\leq \delta_1$, for all $\delta_1\in[0,\breve{a}_k-\hat{a}_k]$ based on the definition of $\upsilon_{\mathcal{H}_0,k}$ and this fact will be used in the proof of Proposition \ref{pro}.

\noindent {\em (Step III).} In this step, we show that there exists a $t_3\in[t_1,t_2+\tau_d]$ such that $x_{ik}(t_3)\leq M_k(x(t_1))-\varepsilon$ and conclude the proof by using Lemma \ref{lem:bound1}.

Define $\varepsilon=\min\{\frac{\gamma\tau_d\delta_1}{L_1^+\tau_d+1},\delta_1\}$ for Assumption \ref{assm-strict} and $\varepsilon=\upsilon_{\mathcal{H}_0,k}(\delta_1)\leq \frac{\tau_d d_{\psi,k}(\delta_1)}{\tau_dL_1^++1}$ for Assumption \ref{assm-hyper}. It follows that $(\gamma\delta_1-L_1^+\varepsilon)\tau_d\geq\varepsilon$ for Assumption \ref{assm-strict} and $(d_{\psi,k}(\delta_1)-L_1^+\varepsilon)\tau_d\geq\varepsilon$ for Assumption \ref{assm-hyper}. Since $\varepsilon>0$, we know that $f^{ik}_{\sigma(t)}(x(t))$ does not change sign and $|f^{ik}_{\sigma(t)}(x(t))|\geq \frac{\varepsilon}{\tau_d}$ for $t\in[t_2,t_2+\tau_d]$. Moreover,
\begin{align}
\left| \|x_{i}(t_2+\tau_d)\|_\mathcal{B}-\|x_{i}(t_2)\|_\mathcal{B}\right|
=\int_{t_2}^{t_2+\tau_d}|D^+\|x_i(\tau)\|_{\mathcal{\mathcal{B}}}|\d\tau\geq\tau_d\frac{\varepsilon}{\tau_d}=\varepsilon.
\end{align}
This contradicts the assumption that $0<\|x_i(t)\|_\mathcal{B}\leq \varepsilon$ for all $t\in[t_1,t_2+\tau_d]$. Thus, there exists a $t_3\in[t_1,t_2+\tau_d]$ such that $x_{ik}(t_3)\leq M_k(x(t_1))-\varepsilon$.

Finally, based on Lemma \ref{lem:bound1}, we obtain $x_{ik}(t)\leq M_k(x(t_1))-\delta_2$ for all $t\in[t_3,T^*]$, where $\delta_2=e^{-L_1^*T^*}\varepsilon$. This completes the proof of the lemma.
\hfill$\square$

The following lemma is symmetric to Lemma \ref{lem:contraction1}.
\begin{lem}\label{lem:contraction2}
Let Assumptions \ref{assm0} and \ref{assm1} hold and assume that $\mathcal{G}_{\sigma(t)}$ is uniformly jointly quasi-strongly connected. Fix any $k\in\mathcal{D}$. For any $(t_1,x(t_1))\in \mathbb{R}\times \mathcal{H}_0^n$, any $\delta_1>0$ and any $T^*>0$, assume that there is an arc $( j, i)$ and a time $t_2\geq t_1$ such that $ j\in \mathcal{N}_i(\sigma(t))$, and $x_{jk}(t)\geq m_k(x(t_1))+\delta_1$. Then, there exists a $t_3\in[t_1,t_2+\tau_d]$ such that $x_{ik}(t)\geq m_k(x(t_1))+\delta_2$, for all $t\in[t_3,t_3+T^*]$. Here, if Assumption \ref{assm-strict} is satisfied, $\delta_2=e^{-L_2^*T^*}\min\{\frac{\gamma\tau_d\delta_1}{L_2^+\tau_d+1},\delta_1\}$ for some positive constants $L_2^*$ and $L_2^+$ related to $\mathcal{H}_0$. If Assumption \ref{assm-hyper} is satisfied, $\delta_2=e^{-L_2^*T^*}\overline{\upsilon}(\delta_1)$ for some positive constant $L_2^*$ and a continuous positive-definite function $\overline{\upsilon}(\cdot)$ both related to $\mathcal{H}_0$.
\end{lem}

\subsection{Proof of Theorem \ref{thm1}}\label{sec:proof1}

The necessity proof follows a similar argument as the proof of Theorem 3.8 of \cite{LinZhiyun_SIAM07}. It is therefore omitted. We focus on the sufficiency and first give an outline of how the lemmas on invariant set, ``interior'' agents, and ``boundary'' agents are used to prove Theorem \ref{thm1}.

The sufficiency proof is outlined as follows.
We first use Lemma \ref{lem:invariant1} to show that point-wise uniform agreement is achieved on $\mathcal{S}_0$. We then focus on agreement attraction.
A common Lyapunov function is constructed and Lemma \ref{lem:invariant1} is used to show that this Lyapunov function is nonincreasing. When the Lyapunov function is not equal to zero initially, we know that there exists at least one agent not on the upper boundary or not on the lower boundary at the initial time. Then, we apply Lemma \ref{lem:bound1} or \ref{lem:bound2} to show that this ``interior'' agent will not become a ``boundary'' agent afterwards. Based on the fact that the interaction graph is uniformly jointly quasi-strongly connected, we show that another agent will be attracted by this ``interior'' agent at a certain time instant. Using Lemma \ref{lem:contraction1} or \ref{lem:contraction2}, we know that this agent will become an ``interior'' agent and will not go back to the boundary. Repeating this process, no agents will stay on the boundary after  certain time. This shows that the Lyapunov function is strictly shrinking, which verifies the desired theorem.

Choose any $\eta\in\mathcal{J}$ and any $\varepsilon>0$, where $\mathcal{J}=\{x\in \mathcal{S}_0^n:x_1=x_2=\dots=x_n\}$. We define $\mathcal{A}_{a}(\eta)=\{x\in\mathcal{S}_0^n:\|x-\eta\|_{\infty}\leq a\}$. It is obvious from Lemma \ref{lem:invariant1} that $\mathcal{A}_{a}(\eta)$ is an invariant set since a hypercube is a special case of a hyperrectangle. Therefore, by setting $\delta=\frac{\varepsilon}{\sqrt{n}}$, we know that
$
\|x(t_0)-\eta\|\leq \delta\quad \Rightarrow\quad  \|x(t)-\eta\|\leq \varepsilon,~~\forall t\geq t_0.
$
This shows that point-wise uniform agreement is achieved on $\mathcal{S}_0$.

Now define
$
V(x)=\rho(\mathcal{H}(\mathcal{C}(x))),
$ where $\rho(\mathcal{H}(\mathcal{C}(x)))$ denotes the maximum side length of the hyperrectangle $\mathcal{H}(\mathcal{C}(x))$. Clearly, it follows from Lemma \ref{lem:invariant1} that $V(x)$ is nonincreasing along \eqref{eq:switch} and $x_i(t)\in\mathcal{H}_0$, $\forall i\in \mathcal{V}$, $\forall t\geq t_0$. We next prove the sufficiency of Theorem \ref{thm1} by showing that $V(x)$ is strictly shrinking over suitable time intervals.

Since $\mathcal{G}_{\sigma(t)}$ is uniformly jointly quasi-strongly connected, there is a $T>0$ such that the union graph $\mathcal{G}([t_0,t_0+T])$ is quasi-strongly connected. Define $T_1=T+2\tau_d$, where $\tau_d$ is the dwell time. Denote
$\kappa_1=t_0+\tau_d$, $\kappa_2=t_0+T_1+\tau_d$, $\dots$, $\kappa_{n^2}=t_0+(n^2-1)T_1+\tau_d$. Thus, there exists a node $i_0\in \mathcal{V}$ such that
$ {i_0}$ has a path to every other nodes jointly on time interval $[\kappa_{l_i},\kappa_{l_i}+T]$, where $i=1,2,\dots,n$ and $1\leq l_1\leq l_2\leq \dots \leq l_n\leq n^2$. Denote $\overline{T}=n^2T_1$.
We divide the rest of the proof into three steps.

\noindent {\em (Step I).} Consider the time interval $[t_0,t_0+\overline{T}]$ and $k=1$. In this step, we show that an agent that does not belong to the interior set will become an ``interior'' agent due to the attraction of ``interior'' agent ${i_0}$.

More specifically, define $\varepsilon_1=\frac{M_1(x(t_0))-m_1(x(t_0))}{2}$.
It is trivial to show that $M_1(x(t)) = m_1(x(t))$, $\forall t\geq t_0$ when $M_1(x(t_0)) = m_1(x(t_0))$ based on Definition \ref{def:cone}. Therefore,
we assume that $M_1(x(t_0))\neq m_1(x(t_0))$ without loss of generality.
Split the node set into two disjoint subsets $\mathcal{V}_1=\{ j|~x_{j1}(t_0)\leq M_1(x(t_0))- \varepsilon_1\}$ and $\overline{\mathcal{V}}_1=\{ j| j\notin\mathcal{V}_1\}$.

Assume that ${i_0}\in\mathcal{V}_1$. This implies that $x_{i_01}(t_0)\leq M_1(x(t_0))- \varepsilon_1$. It follows from Lemma \ref{lem:bound1} that $x_{i_01}(t)\leq M_1(x(t_0))-\delta_1$, $\forall t\in [t_0,t_0+\overline{T}]$, where $\delta_1=e^{-L_1^*\overline{T}}\varepsilon_1$. Considering the time interval $[\kappa_{l_1},\kappa_{l_1}+T]$, we can show that there is an arc $( {i_1}, {j_1})\in \mathcal{V}_1\times \overline{\mathcal{V}}_1$ such that $ {i_1}$ is a neighbor of $ {j_1}$ ($i_1$ might be equal or not to $i_0$) because otherwise there is no arc $( {i_1}, {j_1})$ for any $ {i_1}\in \mathcal{V}_1$ and $ {j_1}\in \overline{\mathcal{V}}_1$
(which contradicts the fact that $ {i_0}\in \mathcal{V}_1$ has a path to every other nodes jointly on time interval $[\kappa_{l_1},\kappa_{l_1}+T]$). Therefore, there exists a time $\tau\in [\kappa_{l_1},\kappa_{l_1}+T]=[t_0+(l_1-1)T+\tau_d,t_0+l_1T-\tau_d]$ such that $ {j_1}\in \mathcal{N}_i(\sigma(\tau))$. Based on Assumption \ref{assm0}, it follows that there is time interval $[\overline{\tau}_1,\overline{\tau}_1+\tau_d]\subset [t_0+(l_1-1)T,t_0+l_1T]$ such that ${j_1}\in \mathcal{N}_i(\sigma(\tau))$, for all $t\in[\overline{\tau}_1,\overline{\tau}_1+\tau_d]$.

Also note that $ {i_1}\in\mathcal{V}_1$ implies that $x_{i_1 1}(t_0)\leq M_1(x(t_0))-\varepsilon_1$. This shows that $x_{i_1 1}(t)\leq M_1(x(t_0))-\delta_1$, $\forall t\in [t_0,t_0+\overline{T}]$ based on Lemma \ref{lem:bound1}. Therefore, it follows from Lemma \ref{lem:contraction1} that there exists a $t_2\in[t_0,\overline{\tau}_1+\tau_d]$ such that
$x_{j_1 1}(t_2)\leq M_1(x(t_0))-\varepsilon_2$ and
$x_{j_1 1}(t)\leq M_1(x(t_0))-\delta_2$, $\forall t\in [t_2,t_2+\overline{T}]$, where $\varepsilon_2=\min\left\{\frac{\gamma\tau_d\delta_1}{L_1^+\tau_d+1},\delta_1\right\}$ and $\delta_2=e^{-L_1^*\overline{T}}\min\left\{\frac{\gamma\tau_d\delta_1}{L_1^+\tau_d+1},\delta_1\right\}$.
To this end, we have shown that at least two agents are not on the upper boundary at $t_0+l_1T$.

\noindent {\em (Step II).}
In this step, we show that the side length of the hyperrectangle $\mathcal{H}(\mathcal{C}(x))$ parallel to the $k\th$ axis $\overrightarrow{r_k}$ at $t_0+\overline{T}$ is strictly less than that at $t_0$.

We can now redefine two disjoint subsets
$\mathcal{V}_2=\{ j|~x_{j1}(t_0)\leq M_1(x(t_0))- \varepsilon_2\}$ and $\overline{\mathcal{V}}_2=\{ j| j\notin\mathcal{V}_2\}$. It then follows that $\mathcal{V}_2$ has at least two nodes by noting that $\varepsilon_2\leq \varepsilon_1$.
By repeating the above analysis, we can show that
$x_{i1}(t)\leq M_1(x(t_0))-\delta_n$, $\forall i\in \mathcal{V}$, $\forall t\in [t_n,t_n+\overline{T}]$ by noting that $\delta_n\leq \delta_{n-1}\leq \dots\leq \delta_1$, where $t_n\in[t_0,\overline{\tau}_n+\tau_d] \subseteq[t_0+(l_n-1)T_1,t_0+l_nT_1]$,
$\delta_n=e^{-nL_1^*\overline{T}}\min\left\{\frac{(\gamma\tau_d)^{n-1}}{(L_1^+\tau_d+1)^{n-1}},1\right\}\varepsilon_1$.

Instead, if $ {i_0}\in\overline{\mathcal{V}}_1$, or what is equivalent, $x_{i_01}(t_0)\geq m_1(x(t_0))+ \varepsilon_1$, we can similarly show that $x_{i1}(t)\geq m_1(x(t_0))+\overline{\delta}_n$, $\forall i\in \mathcal{V}$, $\forall t\in [t_n,t_n+\overline{T}]$ using Lemmas \ref{lem:bound2} and \ref{lem:contraction2}, where $t_n\in[t_0,\overline{\tau}_n+\tau_d] \subseteq[t_0+(l_n-1)T_1,t_0+l_nT_1]$,
$\overline{\delta}_n=e^{-nL_2^*\overline{T}}\min\left\{\frac{(\gamma\tau_d)^{n-1}}{(L_2^+\tau_d+1)^{n-1}},1\right\}\varepsilon_1$.

Therefore, it follows that $D_1(\mathcal{H}(x(t_0+\overline{T})))\leq D_1(\mathcal{H}(x(t_0)))-\beta D_1(\mathcal{H}(x(t_0)))$, and $\beta$ is specified as $\beta=e^{-nL^*\overline{T}}\min\left\{\frac{(\gamma\tau_d)^{n-1}}{2(L^+\tau_d+1)^{n-1}},\frac{1}{2}\right\}$, $L^*=\max\{L_1^*,L_2^*\}$, and $L^+=\max\{L_1^+,L_2^+\}$.

\noindent {\em (Step III)} In this step, we show that $\rho(\mathcal{H}(\mathcal{C}(x)))$ at $t_0+d\overline{T}$ is strictly less than at $t_0$ and thus prove the theorem by showing that $V$ is strictly shrinking.

We consider the time interval $[t_0+\overline{T},t_0+2\overline{T}]$ and $k=2$. Following similar analysis as of Step I and Step II, we can show that $D_2(\mathcal{H}(x(t_0+2\overline{T})))\leq D_2(\mathcal{H}(x(t_0)))-\beta D_2(\mathcal{H}(x(t_0)))$.

By repeating the above analysis, it follows that
$
V(x(t_0+d\overline{T}))-V(x(t_0))\leq -\beta(V(x(t_0))).
$

Then, letting $N$ be the smallest positive integer such that $t\leq t_0+Nd\overline{T}$, we know that
\begin{align}
V(x(t))\leq ~ (1-\beta )^{N-1} V(x(t_0))
 \leq ~\frac{1}{1-\beta}(1-\beta)^{\frac{t-t_0}{d\overline{T}}}V(x(t_0))
 = ~\frac{1}{1-\beta}e^{-\beta^*(t-t_0)}V(x(t_0)),
\end{align}
where $\beta^*=\frac{1}{d\overline{T}}\ln\frac{1}{1-\beta}$. Denote $\mathcal{H}(\mathcal{S}_0)$ as the supporting hyperrectangle of $\mathcal{S}_0$. Since $x(t_0)\in \mathcal{H}_0^n\subseteq\mathcal{H}^n(\mathcal{S}_0)$, it follows  that the above inequality holds for any $x(t_0)\in\mathcal{H}^n(\mathcal{S}_0)$ or any $x(t_0)\in\mathcal{S}_0^n$.
By choosing $k=\frac{1}{1-\beta}$ and $\lambda=\beta^*$, we have that exponential agreement attraction is achieved on $\mathcal{S}_0$. This proves the desired theorem.
\hfill$\square$

\subsection{Proof of Proposition \ref{pro}}\label{sec:proof1-2}

The necessity proof follows a similar argument as the proof of Theorem 3.8 of \cite{LinZhiyun_SIAM07} and the proof of point-wise uniform agreement is similar to the one of Theorem \ref{thm1}. We focus on
the proof of agreement attraction and use a similar analysis as  in the proof of Theorem \ref{thm1}.

Using the same Lyapunov function
$
V(x)=\rho(\mathcal{H}(\mathcal{C}(x)))
$ as in the proof of Theorem \ref{thm1}, we first show that
$x_{i_01}(t_0)\leq M_1(x(t_0))- \varepsilon_1$ and $x_{i_01}(t)\leq M_1(x(t_0))-\delta_1$, $\forall t\in [t_0,t_0+\overline{T}]$, where $\varepsilon_1=\frac{M_1(x(t_0))-m_1(x(t_0))}{2}$ and $\delta_1=e^{-L_1^*\overline{T}}\varepsilon_1$. Then, we have another agent
$ {i_1}\in\mathcal{V}_1$ satisfying $x_{i_1 1}(t_0)\leq M_1(x(t_0))-\varepsilon_1$. This shows that $x_{i_1 1}(t)\leq M_1(x(t_0))-\delta_1$, $\forall t\in [t_0,t_0+\overline{T}]$ based on Lemma \ref{lem:bound1}. Therefore, it follows from Lemma \ref{lem:contraction1} that there exists $t_2\in[t_0,\overline{\tau}_1+\tau_d]$ such that
$x_{j_1 1}(t_2)\leq M_1(x(t_0))-\varepsilon_2$ and
$x_{j_1 1}(t)\leq M_1(x(t_0))-\delta_2$, $\forall t\in [t_2,t_2+\overline{T}]$, where $\varepsilon_2=\upsilon_{\mathcal{H}_0,k}(\delta_1)$ and $\delta_2=e^{-L_1^*\overline{T}}\upsilon_{\mathcal{H}_0,k}(\delta_1)$.

Then, we define two disjoint subsets
$\mathcal{V}_2=\{ j|~x_{j1}(t_0)\leq M_1(x(t_0))- \varepsilon_2\}$ and $\overline{\mathcal{V}}_2=\{ j| j\notin\mathcal{V}_2\}$. It follows that $\mathcal{V}_2$ has at least two nodes. Note that $\upsilon_{\mathcal{H}_0,k}(\delta_1)\leq \delta_1$ for all its definition domain.
By repeating the above analysis, we can show that
$x_{i1}(t)\leq M_1(x(t_0))-\delta_n$, $\forall i\in \mathcal{V}$, $\forall t\in [t_n,t_n+\overline{T}]$ by noting that $\delta_n\leq \delta_{n-1}\leq \dots\leq \delta_1$, where $t_n\in[t_0,\overline{\tau}_n+\tau_d] \subseteq[t_0+(l_n-1)T_1,t_0+l_nT_1]$, $\delta_n(D_1(\mathcal{H}(x(t_0))))=\varsigma\circ\upsilon_{\mathcal{H}_0,k}\circ\dots\upsilon_{\mathcal{H}_0,k}\circ\varsigma(\frac{\cdot}{2})
$, a continuous positive-definite function $\varsigma(\cdot)$ is defined as $\varsigma(x)=e^{-L_1^*\overline{T}}x$, and $D_1(\mathcal{H}(x(t_0)))=M_1(x(t_0))-m_1(x(t_0))$. It is obvious that $\delta_n(D_1(\mathcal{H}(x(t_0))))$ is a continuous positive-definite function.

Instead, if $ {i_0}\in\overline{\mathcal{V}}_1$, or what is equivalent, $x_{i_01}(t_0)\geq m_1(x(t_0))+ \varepsilon_1$, we can similarly show that $x_{i1}(t)\geq m_1(x(t_0))+\overline{\delta}_n$, $\forall i\in \mathcal{V}$, $\forall t\in [t_n,t_n+\overline{T}]$ using Lemmas \ref{lem:bound2} and \ref{lem:contraction2}, where $t_n\in[t_0,\overline{\tau}_n+\tau_d] \subseteq[t_0+(l_n-1)T_1,t_0+l_nT_1]$,
a continuous positive-definite function $\overline{\delta}_n(D_1(\mathcal{H}(x(t_0))))=\overline{\varsigma}\circ\overline{\upsilon}_{\mathcal{H}_0}\circ\dots
\overline{\upsilon}_{\mathcal{H}_0}\circ\overline{\varsigma}(\frac{\cdot}{2})
$, and $\overline{\varsigma}(\cdot)$ is defined as $\varsigma(x)=e^{-L_2^*\overline{T}}x$.
Therefore, it follows that $D_1(\mathcal{H}(x(t_0+\overline{T})))\leq D_1(\mathcal{H}(x(t_0)))-\delta^*(D_1(\mathcal{H}(x(t_0))))$, where $\delta^*(x)=\min\{\delta_n(x),\overline{\delta}_n(x)\}$ is a continuous positive-definite function.

Then, following Lemma 4.3 of \cite{Khalil_book}, there exists a class $\mathcal{K}$ function $\Upsilon(D_1(\mathcal{H}(x(t_0))))$ defined on $[0,\breve{a}_k-\hat{a}_k]$ satisfying $\Upsilon(D_1(\mathcal{H}(x(t_0))))\leq \delta^*(D_1(\mathcal{H}(x(t_0))))$, $\forall D_1(\mathcal{H}(x(t_0)))\in [0,\breve{a}_k-\hat{a}_k]$, where a continuous function $\Upsilon:[0,b)\rightarrow[0,\infty)$ is said to belong to class $\mathcal{K}$ if it is strictly increasing and $\Upsilon(0)=0$. Therefore, it follows that $D_1(\mathcal{H}(x(t_0+\overline{T})))\leq D_1(\mathcal{H}(x(t_0)))-\Upsilon(D_1(\mathcal{H}(x(t_0))))$, $\forall i\in \mathcal{V}$, $\forall t\in [t_n,t_n+\overline{T}]$.

We next consider the time interval $[t_0+\overline{T},t_0+2\overline{T}]$. Following the previous analysis, we can show that $D_2(\mathcal{H}(x(t_0+2\overline{T})))\leq D_2(\mathcal{H}(x(t_0)))-\Upsilon(D_2(\mathcal{H}(x(t_0))))$.

By repeating the above analysis, it follows that
$
V(x(t_0+d\overline{T}))-V(x(t_0))\leq -\Upsilon(V(x(t_0))).
$

Then,
let $N$ be the smallest positive integer such that $t\leq t_0+Nd\overline{T}$. It then follows that
\begin{align}
V(x(t))-V(x(t_0))\leq ~ -\Upsilon(V(x(t_0)))-\dots-\Upsilon(V(x(t_0+(N-1)d\overline{T})))
 \leq  -N\Upsilon(V(x(t_0))).
\end{align}
Therefore, for any $\varepsilon>0$, there exists a sufficiently large $N$ such that
$
V(x(t))\leq\frac{2\varepsilon}{\sqrt{n}},~~\forall t\geq t_0+Nd\overline{T}.
$
This shows that there exists $\eta\in\mathcal{J}$ such that $\|x(t)-\eta\|_{\infty}\leq\frac{\varepsilon}{\sqrt{n}}$, $\forall t\geq t_0+Nd\overline{T}$. This implies $\|x(t)-\eta\|\leq\varepsilon$, $\forall t\geq t_0+Nd\overline{T}$, which shows that uniformly agreement attraction is achieved on $\mathcal{S}_0$ and proves the proposition.
\hfill$\square$

\begin{remark} [Extension to global convergence]
The convergence is semi-global since the selections of $\mathcal{K}$ class function $\Upsilon$, and parameters $\lambda$ and $k$ depend on that the initial common space is given in advance and compact, i.e., the assumption that $\mathcal{S}_0$ is compact is necessary to guarantee uniformly asymptotic or exponential agreement. On the other hand, if Assumption \ref{assm1} is changed to ``uniformly globally Lipschitz'', we obtain a global convergence result.
\end{remark}

\section{Cooperative--antagonistic Multi-agent Systems}\label{sec:antagonistic}

In this section, we focus on \\ cooperative--antagonistic multi-agent systems and prove Theorem~\ref{thm2} using a contradiction argument, with the help of a
series of preliminary lemmas. Note that since every agent admits a continuous trajectory, we only need to prove that all the agents' componentwise absolute values reach an agreement.

\subsection{Invariant set}\label{sec:invariant2}

In this section, we construct an invariant set for the dynamics under the cooperative--antagonistic network. For all $k\in \mathcal{D}$,
define
$
M^\dag_k(x(t))=\max_{i\in \mathcal{V}}|x_{ik}(t)|.
$
In addition, define an origin-symmetric supporting hyperrectangle $\mathcal{H}(\widehat{\mathcal{C}}(x))\subset \mathbb{R}^d$ as
$
\mathcal{H}(\widehat{\mathcal{C}}(x)):=[-M^\dag_1(x),M^\dag_1(x)]\times\dots\times[-M^\dag_d(x),M^\dag_d(x)].
$
The origin-symmetric supporting hyperrectangle formed by the initial states of all agents is given by $\widehat{\mathcal{H}}_0^n$, where
\begin{align}
\widehat{\mathcal{H}}_0=\big[-\max_{i\in \mathcal{V}}|x_{i1}(t_0)|,\max_{ i\in \mathcal{V}}|x_{i1}(t_0)|\big]\times\dots\times \big[-\max_{i\in \mathcal{V}}|x_{id}(t_0)|,\max_{i\in \mathcal{V}}|x_{id}(t_0)|\big].
\end{align}

Introduce the state transformation
$
y_{ik}=x_{ik}^2,
$
for all $i\in \mathcal{V}$, and for all $k\in \mathcal{D}$.
The analysis will be carried out on $y_{ik}$, instead of $x_{ik}$ to avoid non-smoothness.
The following lemma establishes an invariant set for system \eqref{eq:switch}.
\begin{lem}\label{lem:invariant2}
Let Assumptions \ref{assm0}, \ref{assm1} and \ref{assm3} hold. Then, for system \eqref{eq:switch}, $\widehat{\mathcal{H}}_0^n$ is an invariant set, i.e., $x_i(t)\in \widehat{\mathcal{H}}_0$, $\forall i\in \mathcal{V}$, $\forall t\geq t_0$.
\end{lem}

{\it Proof.}
Let $y_k=\max_{i\in\mathcal{V}}y_{ik}$, for all $k\in \mathcal{D}$.
We first show that $D^+y_k\leq 0$, for all $k\in \mathcal{D}$.
It follows from \eqref{eq:set-deri} that
$
\dot y_{ik}=2x_{ik}f_{\sigma(t)}^{ik}(x),~~\forall i\in \mathcal{V},~ \forall k\in \mathcal{D}.
$
Let $\widehat{\mathcal{V}}(t)=\{i\in\mathcal{V}: y_{ik}(x_i(t))=y_k(x(t))\}$ be the set of indices where the maximum is reached at $t$.
It then follows from Lemma \ref{lem:Dini} that for all $k\in \mathcal{D}$,
$
D^+y_{k}=2\max_{i\in\widehat{\mathcal{V}}(t)} x_{ik}f_{\sigma(t)}^{ik}(x).
$
Consider any $x(t_0)\in\mathcal{H}_0^n$ and any initial time $t_0$.
It follows from Definition \ref{def:cone} and Lemma \ref{lem:cone} that
$
f_p^i(x)\in\mathcal{T}_\gamma(x_i,\mathcal{H}(\overline{\mathcal{C}}_p^i(x)))\subseteq\mathcal{T}(x_i,\mathcal{H}
(\widehat{\mathcal{C}}(x))),
\quad \forall i\in \mathcal{V}, ~\forall p\in \mathfrak{P}.
$
Based on Definition \ref{def:cone}, it follows that $ f_p^{ik}(x)\leq 0$ for $x_{ik}=\sqrt{y_k}\geq 0$ and $ f_p^{ik}(x)\geq 0$ for $x_{ik}=-\sqrt{y_k}\leq 0$. This shows that $x_{ik}f_p^{ik}(x)\leq 0$ for  $i\in\widehat{\mathcal{V}}=\{i\in\mathcal{V}: y_{ik}=y_k\}$. It follows that for all $k\in \mathcal{D}$, and $x\in \widehat{\mathcal{H}}_0^n$,
$
D^+y_{k}\leq 0.
$
Therefore, $x_{ik}^2(t)\leq \max_{i\in \mathcal{V}}x_{ik}^2(t_0)$, $\forall i\in \mathcal{V}$, $\forall k\in \mathcal{D}$, which shows that
$
-\max_{i\in \mathcal{V}}|x_{ik}(t_0)|\leq x_{ik}(t)\leq \max_{i\in \mathcal{V}}|x_{ik}(t_0)|,~\forall i\in \mathcal{V},~\forall k\in \mathcal{D},~\forall t\geq t_0.
$
This implies that $\widehat{\mathcal{H}}_0^n$ is an invariant set.
\hfill$\square$


\subsection{``Interior'' agents}\label{sec:interior2}

In the following lemma, we show that the projection of the state on any axis is strictly less
than a certain upper bound as long as it is initially strictly less than this upper bound. The lemma relies on the technical Lemmas \ref{lem:Lipschitz1-1} and \ref{lem:Lipschitz2-1}, which can be found in the appendices.

\begin{lem}\label{lem:bound3}
Let Assumptions \ref{assm0}, \ref{assm1}, and \ref{assm3} hold and assume that $\mathcal{G}_{\sigma(t)}$ is uniformly jointly strongly connected. Fix any $k\in\mathcal{D}$. For any $(t_1,x(t_1))\in \mathbb{R}\times \widehat{\mathcal{H}}_0^n$, any $\varepsilon>0$ and any $T^*>0$, if $y_{ik}(t_2)\leq y^* -\varepsilon$ at some $t_2\geq t_1$, where $y^*\geq y_k(x(t_1))$ is a constant, then $y_{ik}(t)\leq y^*-\delta$ for all $t\in[t_2,t_2+T^*]$, where $\delta=e^{-L^*T^*}\varepsilon$, and $L^*$ is a positive constant related to $\widehat{\mathcal{H}}_0$.
\end{lem}

{\it Proof.}
Similar to the proof of Lemma \ref{lem:bound1}, we consider any $(t_1,x(t_1))\in \mathbb{R}\times \widehat{\mathcal{H}}_0^n$, and let $\psi=x(t_1)$ and  $\mathcal{M}_{ik}=\widehat{\mathcal{H}}(\psi)\times\dots\times\widehat{\mathcal{H}}(\psi)\times\widehat{\mathcal{H}}^o_k(\psi)
\times\widehat{\mathcal{H}}(\psi)\times\dots\times \widehat{\mathcal{H}}(\psi)$, where $\widehat{\mathcal{H}}(\psi)=[-M^\dag_1(\psi),M^\dag_1(\psi)]\times\dots \times[-M^\dag_d(\psi),M^\dag_d(\psi)]$, and $\widehat{\mathcal{H}}^o_k(\psi)=[-M^\dag_1(\psi),M^\dag_1(\psi)]\times\dots \\ \times [-M^\dag_{k-1}(\psi),M^\dag_{k-1}(\psi)]
\times[-M^\dag_{k+1}(\psi),M^\dag_{k+1}(\psi)]\times\dots  \times[-M^\dag_{d}(\psi),M^\dag_{d}(\psi)]$. Again, for clarity we divide the rest of the proof into three steps.

\noindent{ \it (Step I).}
Define the following function
\begin{align}
g_{\psi,k}(\chi):[-M^\dag_k(\psi), M^\dag_k(\psi)]\rightarrow \mathbb{R}, ~~
\chi\rightarrow
\sup_{p\in\mathfrak{P}}\{\max_{i\in\mathcal{V}}\{\max_{y\in \mathcal{M}_{ik}}\{ x_{ik}f_p^{ik}(x_{ik},y):x_{ik}=\chi\}  \}\}.
\end{align}
Obviously, $g_{\psi,k}$ is continuous. In this step, we establish some useful properties of $g_{\psi,k}$ based on Lemmas \ref{lem:Lipschitz1-1} and \ref{lem:Lipschitz2-1}. We make the following claim.

Claim A: (i) $g_{\psi,k}(\chi)=0$ if $\chi=\pm M^\dag_k(\psi)$; (ii) $g_{\psi,k}(\chi)>0$ if $\chi \in (-M^\dag_k(\psi),M^\dag_k(\psi))$; (iii) $g_{\psi,k}(\chi)$ is Lipschitz continuous with respect to $\chi$ on $[-M^\dag_k(\psi), M^\dag_k(\psi)]$.

The first and second properties of Claim A can be obtained following a similar analysis to the proof of Lemma \ref{lem:bound1}.
For the third property of Claim A, it follows from Lemma \ref{lem:Lipschitz2-1} that $g_p^{ik}:[-M^\dag_k(\psi), M^\dag_k(\psi)]\rightarrow \mathbb{R}$, $x_{ik}\mapsto\max_{y\in \mathcal{M}_{ik}} x_{ik}f_p^{ik}(x_{ik},y)$ is Lipschitz continuous with respect to $x_{ik}$, $\forall k\in\mathcal{D}$, $\forall i\in\mathcal{V}$, and $\forall p\in\mathfrak{P}$. Also note that $g_{\psi,k}(M^\dag_k(\psi))=0$. Then, it follows from Lemma \ref{lem:Lipschitz1-1} that $g_{\psi,k}(\chi)$ is Lipschitz continuous with respect to $\chi$ on $[-M^\dag_k(\psi), M^\dag_k(\psi)]$.
\vspace{2mm}

\noindent{ \it (Step II).} In this step, we construct another nonlinear function $h_{\widehat{\mathcal{H}}_0,k}(\cdot)$ based on the definition of $g_{\psi,k}(\cdot)$.
From the definitions of $g_{\psi,k}(\chi)$, it follows that
$
\dot y_{ik}(t)=2 x_{ik}f_{\sigma(t)}^{ik}(x(t))\leq 2g_{\psi,k}(x_{ik}(t)),~ \forall t\geq t_2.
$
It also follows from the properties of $g_{\psi,k}(\chi)$ that there exists a Lipschitz constant $L_1$ such that $g_{\psi,k}(\chi)=|g_{\psi,k}(\chi)-g_{\psi,k}(M^\dag_{k}(\psi))|\leq L_1|\chi-M^\dag_k(\psi)|=L_1(M^\dag_k(\psi)-\chi)$, $\forall \chi\in [-M^\dag_k(\psi), M^\dag_k(\psi)]$ and $g_{\psi,k}(\chi)=|g_{\psi,k}(\chi)-g_{\psi,k}(-M^\dag_{k}(\psi))|\leq L_1|\chi+M^\dag_k(\psi)|=L_1(M^\dag_k(\psi)+\chi)$, $\forall \chi\in [-M^\dag_k(\psi), M^\dag_k(\psi)]$, where $L_1$ is related to $\psi$.

Therefore, for the case of $x_{ik}\geq 0$, we have that
$\dot y_{ik}(t)\leq 2L_1(M^\dag_k(\psi)-x_{ik})=2L_1(M^\dag_k(\psi)-\sqrt{y_{ik}(t)})$,
For the case of $x_{ik}<0$, we have that $\dot y_{ik}(t)\leq 2L_1(M^\dag_k(\psi)+x_{ik})=2L_1(M^\dag_k(\psi)-\sqrt{y_{ik}(t)})$. Overall, it follows that
$
\dot y_{ik}(t)\leq 2L_1(M^\dag_k(\psi)-\sqrt{y_{ik}(t)}),~ \forall t\geq t_2.
$
Let $\phi(t)$ be the solution of $\dot\phi=\overline{h}_{\psi,k}(\phi)$ with initial condition $\phi(t_2)=y_{ik}(t_2)$, where
$\overline{h}_{\psi,k}:[0,y_k(\psi)]\rightarrow \mathbb{R}$, $\phi\mapsto 2L_1(M^\dag_k(\psi)-\sqrt{\phi})$.
It follows from the Comparison Lemma that $y_{ik}(t)\leq \phi(t)$, $\forall t\geq t_2$.

Next, by letting $\varphi=y_{k}(\psi)-\phi$ and $\breve{a}_k=M^\dag_k(x(t_0))$, we define the following function
\begin{align}
h_{\widehat{\mathcal{H}}_0,k}(\varphi):[0,(\breve{a}_k)^2] \rightarrow \mathbb{R},~~
\varphi\mapsto\begin{cases}
\overline{h}_{\psi,k}(y_{k}(\psi)-\varphi); \quad \text{if}~~ \varphi\in[0, y_k(\psi)],\\
\overline{h}_{\psi,k}(0); \quad \text{if}~~\varphi\in(y_k(\psi),(\breve{a}_k)^2].\\
\end{cases}
 \end{align}
We have the following claim by easily checking the definition of $h_{\widehat{\mathcal{H}}_0,k}(\varphi)$.

Claim B: (i) $h_{\widehat{\mathcal{H}}_0,k}(\varphi)$ is Lipschitz continuous with respect to $\varphi$ on $[0,(\breve{a}_k)^2]$; (ii) $h_{\widehat{\mathcal{H}}_0,k}(\varphi)=0$ if $\varphi=0$; (iii) $h_{\widehat{\mathcal{H}}_0,k}(\varphi)> 0$ if $\varphi\in (0,(\breve{a}_k)^2]$.

\noindent{ \it (Step III).}
In this step, we take advantage of the function $h_{\widehat{\mathcal{H}}_0,k}(\cdot)$ to show that $y_{ik}$ will be always strictly less than the upper bound $y^*$ as long as it is initially strictly less than $y^*$.

Consider any $T^*>0$ and $t\in[t_2,t_2+T^*]$. It follows from the first property of $h_{\widehat{\mathcal{H}}_0,k}(\varphi)$ that there exists a constant $L^*$ related to $\widehat{\mathcal{H}}_0$ such that $|h_{\widehat{\mathcal{H}}_0,k}(\varphi)-h_{\widehat{\mathcal{H}}_0,k}(0)|\leq L^*\varphi$, $\forall \varphi\in[0,(\breve{a}_k)^2]$. From the second and third properties of $h_{\widehat{\mathcal{H}}_0,k}(\varphi)$, it follows that $h_{\widehat{\mathcal{H}}_0,k}(\varphi)\leq L^*\varphi$, $\forall \varphi\in[0,(\breve{a}_k)^2]$ and  $\dot\varphi=-\dot\phi=-\overline{h}_{\psi,k}(\phi)=-h_{\widehat{\mathcal{H}}_0,k}(\varphi)$, $\forall \varphi\in[0, y_k(\psi)]$. It follows from the Comparison Lemma that
the solution of $\dot\varphi=-h_{\widehat{\mathcal{H}}_0,k}(\varphi)$ satisfies $\varphi(t)\geq e^{-L^*(t-t_2)}\varphi(t_2)$, $\forall t\geq t_2$ since $-h_{\widehat{\mathcal{H}}_0,k}(\varphi)\geq -L^*\varphi$.

Therefore, $y_{ik}(t)\leq \phi(t)=y_{k}(\psi)-\varphi(t)\leq y_{k}(\psi)-e^{-L^*(t-t_2)}(y_{k}(\psi)-\phi(t_2))
\leq y_{k}(\psi)-e^{-L^*T^*}(y_{k}(\psi)-y_{ik}(t_2))=y^*+y_k(\psi)-y^*-e^{-L^*T^*}(y^*+y_k(\psi)-y^*-y_{ik}(t_2))
=y^*-e^{-L^*T^*}(y^*-y_{ik}(t_2)) +(y_k(\psi)-y^*)(1-e^{-L^*T^*}) \leq y^*-e^{-L^*T^*}\varepsilon$ for all $t\in[t_2,t_2+T^*]$ since $y^*\geq y_k(\psi)$. This proves the lemma by letting $\delta=e^{-L^*T^*}\varepsilon$.
\hfill$\square$

\subsection{``Boundary'' agents}\label{sec:boundary2}

In the following lemma, we show that any agent that is attracted by an agent strictly inside the upper
bound is drawn strictly inside that bound. This lemma relies on Lemma \ref{lem:Lipschitz3}, which can be found in the Appendices.
\begin{lem}\label{lem:contraction3}
Let Assumptions \ref{assm0}, \ref{assm1}, and \ref{assm3} hold and assume that $\mathcal{G}_{\sigma(t)}$ is uniformly jointly strongly connected. Fix any $k\in\mathcal{D}$.
For any $(t_1,x(t_1))\in \mathbb{R}\times \widehat{\mathcal{H}}_0^n$ and any $\delta>0$, assume that there is an arc $( j, i)$ and a time $t_2\geq t_1$ such that $ j\in \mathcal{N}_i(\sigma(t))$, and $y_{jk}(t)\leq y^*-\delta$ for all $t\in[t_2,t_2+\tau_d]$, where $y^*\geq y_k(x(t_1))$ is a constant. Then, there exists a $t_3\in[t_1,t_2+\tau_d]$ such that $y_{ik}(t_3)\leq y^*-\varepsilon$, where $\varepsilon=\frac{\gamma\tau_d\delta}{2(L^+\tau_d+\gamma\tau_d+1)}$ and $L^+$ is a constant related to $\widehat{\mathcal{H}}_0$.
\end{lem}

{\it Proof.}
We prove this lemma by contradiction.
Suppose that there does not exist a $t_3\in[t_1,t_2+\tau_d]$ such that $|x_{ik}(t_3)|\leq \sqrt{y^*}-\varepsilon_1$, where $\varepsilon_1=\frac{\gamma\tau_d\delta}{2(L^+\tau_d+\gamma\tau_d+1)\sqrt{y^*}}$ is a positive constant. Then it follows that $\sqrt{y^*}-\varepsilon_1<|x_{ik}(t)|\leq M^\dag_k(x(t_1))$ for all $t\in[t_1,t_2+\tau_d]$.

We focus on the time interval $t\in[t_2,t_2+\tau_d]$.
Define $\overline{x}(t)$ by replacing $x_{ik}(t)$ in $x(t)$ with $\overline{x}_{ik}(t)=\max_{i\in\mathcal{V}}\{x_{ik}(t)\}=M^\dag_k(x(t))$.
We know that $f_p^{ik}(x(t))$ is uniformly locally Lipschitz with respect to $x$ and compact on $\widehat{\mathcal{H}}_0^n$, for all $i\in \mathcal{V}$, and all $p\in\mathfrak{P}$.
By noting that
$M^\dag_k(x(t_1))-\varepsilon_1\leq \sqrt{y^*}-\varepsilon_1<|x_{ik}(t)|$, it follows that there exists a positive constant
$L^+$ related to $\widehat{\mathcal{H}}_0$ such that
$|f_p^{ik}(\overline{x})|-|f_p^{ik}(x)|\leq |f_p^{ik}(\overline{x})-f_p^{ik}(x)|\leq L^+\|x(t)-\overline{x}(t)\|\leq L^+\varepsilon_1$, $\forall p\in\mathfrak{P}$, and $\forall x,\overline{x}\in \widehat{\mathcal{H}}_0^n$.

It follows from $y_{jk}(t)\leq y^*-\delta$ that $\sqrt{y^*}-|x_{jk}(t)|\geq \frac{\delta}{2\sqrt{y^*}}$. Therefore, for any $p^*\in \mathfrak{P}$ such that there is an arc $(j, i)$ with $ j\in \mathcal{N}_i(p^*)$ and $\sqrt{y^*}-|x_{jk}(t)|\geq \frac{\delta}{2\sqrt{y^*}}$,
it follows from Assumption \ref{assm3} that
\begin{align}
\left|f_{p^*}^{ik}(\overline{x})\right|
 \geq  \gamma D_k (\mathcal{H}(\co\{\overline{x}_i,x_j\sgn^{ij}_{p^*}:j\in \mathcal{N}_i(p^*)\}))
 \geq \gamma D_k (\mathcal{H}(\co\{\overline{x}_i,x_j\sgn^{ij}_{p^*}\}))
 > \gamma\left( \frac{\delta}{2\sqrt{y^*}}-\varepsilon_1\right).
\end{align}
Note that $ \frac{\delta}{2\sqrt{y^*}}>\varepsilon_1$ based on the definition of $\varepsilon_1$. Therefore,
we know that $|f_{p^*}^{ik}(x(t))|\geq |f_{p^*}^{ik}(\overline{x}(t))|-L^+\varepsilon_1>\gamma\left( \frac{\delta}{2\sqrt{y^*}}-\varepsilon_1\right)-L^+\varepsilon_1$.
It then follows that for all $t\in [t_2,t_2+\tau_d]$,
$
\left|D^+|x_{ik}(t)|\right|=|f_{\sigma(t)}^{ik}(x(t))|>\gamma\left( \frac{\delta}{2\sqrt{y^*}}-\varepsilon_1\right)-L^+\varepsilon_1.
$
Choose $\varepsilon_1$ sufficiently small, especially, $\varepsilon_1=\frac{\gamma\tau_d\delta}{2(L^+\tau_d+\gamma\tau_d+1)\sqrt{y^*}}$. Such $\varepsilon_1$ exists for every $y^*>0$.
It follows that $\gamma\left( \frac{\delta}{2\sqrt{y^*}}-\varepsilon_1\right)-L^+\varepsilon_1>\frac{\varepsilon_1}{\tau_d}$. Therefore, we know that
\begin{align}
\left | x_{ik}(t_2+\tau_d)-x_{ik}(t_2)\right|
=\int_{t_2}^{t_2+\tau_d}\left|D^+|x_{ik}(\tau)|\right|\d\tau>\tau_d\frac{\varepsilon_1}{\tau_d}=\varepsilon_1.
\end{align}
This contradicts the assumption that $\sqrt{y^*}-\varepsilon_1<|x_{ik}(t)|\leq M^\dag_{k}(x(t_1))$ for all $t\in[t_2,t_2+\tau_d]$.
 It then follows that $y_{ik}(t_3)\leq y^*-\varepsilon$, where $\varepsilon=\sqrt{y^*}\varepsilon_1=\frac{\gamma\tau_d\delta}{2(L^+\tau_d+\gamma\tau_d+1)}$.
\hfill$\square$

\subsection{Proof of Theorem \ref{thm2}}

Unlike the contraction analysis of a common Lyapunov function given in the proof of Theorem \ref{thm1}, we use a contradiction argument for the proof of Theorem \ref{thm2}.

According to the proof of Lemma \ref{lem:invariant2}, we know that $D^+y_k\leq 0$ and $y_k\geq 0$, for all $k\in\mathcal{D}$. Therefore, $y_k(t)$, $k\in\mathcal{D}$ is monotonically decreasing and bounded from below by zero. This implies that for any initial time $t_0$ and initial state $x(t_0)$, there exists a constant $y_k^*$, $k\in \mathcal{D}$ such that
$
 \lim_{t\rightarrow \infty}y_k(t)=y_k^*, ~  \forall k\in \mathcal{D}.
$
Define $\hbar_{ik}=\limsup_{t\rightarrow \infty} y_{ik}(t)$ and $\ell_{ik}=\liminf _{t\rightarrow \infty}y_{ik}(t)$, for all $i\in \mathcal{V}$, and $k\in \mathcal{D}$. Clearly, $0\leq \ell_{ik}\leq \hbar_{ik}\leq y_k^*$.
We know that the componentwise absolute values of all the agents converges to the same if and only if $\hbar_{ik}=\ell_{ik}=y_k^*$, $\forall i\in \mathcal{V}$, $\forall k\in \mathcal{D}$. The desired conclusion holds trivially if $y^*_k=0,  k\in \mathcal{D}$. Therefore, we assume that $y^*_k>0$ for some $k\in \mathcal{D}$ without loss of generality.

Suppose that there exists a node $ {i_1}\in \mathcal{V}$ such that $0\leq \ell_{i_1k}< \hbar_{i_1k}\leq y_k^*$. Based on the fact that $\lim_{t\rightarrow \infty}y_k(t)= y_k^*$, it follows that for any $\overline{\varepsilon}>0$, there exists a $\widehat{t}(\overline{\varepsilon})>t_0$ such that
$
y^*_k-\overline{\varepsilon}\leq y_{k}(t)\leq y^*_k+\overline{\varepsilon}, ~~t\geq \widehat{t}(\overline{\varepsilon}).
$
Take $\alpha_{1k}=\sqrt{\frac{1}{2}(\ell_{i_1k}+\hbar_{i_1k})}$. Therefore, there exists a time $t_1\geq \widehat{t}(\overline{\varepsilon})$ such that $|x_{i_1k}(t_1)|=\alpha_{1k}$ based on the definitions of $\hbar_{i_1k}$ and $\ell_{i_1k}$ and continuousness of $x_{i_1k}(t)$. This shows that
\begin{align}
x_{i_1k}^2(t_1)=\hbar_{i_1k}-(\hbar_{i_1k}-\alpha^2_{1k}) \leq y^*_k+\overline{\varepsilon}-(\hbar_{i_1k}-\alpha^2_{1k})=y^*_k+\overline{\varepsilon}-\varepsilon_1,
\end{align}
where $\varepsilon_1=\hbar_{i_1k}-\alpha^2_{1k}>0$ and the first inequality is based on the definition of $\hbar_{i_1k}$.

Since $\mathcal{G}_{\sigma(t)}$ is uniformly jointly strongly connected, there is a $T>0$ such that the union graph $\mathcal{G}([t_1,t_1+T])$ is jointly strongly connected. Define $T_1=T+2\tau_d$, where $\tau_d$ is the dwell time. Denote
$\kappa_1=t_1+\tau_d$, $\kappa_2=t_1+T_1+\tau_d$, $\dots$, $\kappa_{n}=t_1+(n-1)T_1+\tau_d$. For each node $i\in \mathcal{V}$, $i$ has a path to every other nodes jointly on time interval $[\kappa_{l},\kappa_{l}+T]$, where $l=1,2,\dots,n$.
Denote $\overline{T}=nT_1$.

Consider time interval $[t_1,t_1+\overline{T}]$. Based on the fact that $y_k(x(t_1))\leq y^*_k+\overline{\varepsilon}$ and considering $y^*_k+\overline{\varepsilon}$ as the role of $y^*$ in Lemma \ref{lem:bound3}, it follows that $y_{ik}(t)\leq y^*_k+\overline{\varepsilon}-\delta_1$, $\forall t\in [t_1,t_1+\overline{T}]$, where $\delta_1=e^{-L^*\overline{T}}\varepsilon_1$.

Since for each node $i\in \mathcal{V}$, $i$ has a path to every other nodes jointly on time interval $[\kappa_{l},\kappa_{l}+T]$, where $l=1,2,\dots,n$, there exists $i_2\in \mathcal{V}$ such that $ {i_1}$ is a neighbor of $ {i_2}$ during the time interval $[\kappa_{1},\kappa_{1}+T]$. Based on Lemma \ref{lem:contraction3},
it follows that there exists $t_2\in[t_1,\overline{\tau}_1+\tau_d]\subset[t_1+T,t_1+2T]$ such that $x_{i_2k}^2(t_2)\leq y^*_k+\overline{\varepsilon}-\varepsilon_2$,
where $\varepsilon_2=\frac{\gamma\tau_d}{2(L^+\tau_d+\gamma\tau_d+1)}\delta_1$.
This further implies that $x_{i_2k}^2(t)\leq y^*_k+\overline{\varepsilon}-\delta_2$, $\forall t\in [t_2,t_1+\overline{T}]$, where $\delta_2=e^{-L^*\overline{T}}\varepsilon_2$. By repeating the above analysis, we can show that $y_{ik}(t)\leq y^*_k+\overline{\varepsilon}-\delta_n$, $\forall t\in [t_n,t_1+\overline{T}]$, $\forall i\in \mathcal{V}$, where $t_n\in[t_1,\overline{\tau}_n+\tau_d]\subset[t_1+(n-1)T,t_1+nT]$, and $\delta_n$ can be iteratively obtained as    $\delta_n=e^{-nL^*\overline{T}}\frac{\gamma^{n-1}\tau_d^{n-1}}{2^{n-1}(L^+\tau_d+\gamma\tau_d+1)^{n-1}}$. This is indeed true because $\delta_i\leq \delta_{i-1}$, $\forall i=2,3,\dots,n$.

This shows that $y_k(t_1+\overline{T})=\max_{i\in \mathcal{V}}y_{ik}\leq y_k^*+\overline{\varepsilon}-\delta_n$, which indicates a contradiction for sufficiently small $\overline{\varepsilon}$ satisfying $\overline{\varepsilon}<\delta_n/2$.
Therefore, $\hbar_{ik}=\ell_{ik}=y_k^*$, $\forall i\in \mathcal{V}$, $\forall k\in \mathcal{D}$. This proves that $\lim_{t\rightarrow\infty}(|x_{ik}(t)|-\sqrt{y_k^*})=0$, $\forall i\in \mathcal{V}$ and $\forall k\in \mathcal{D}$, which shows the componentwise absolute values of all the agents converges to the same values and the theorem holds.
\hfill$\square$


\section{Conclusions}\label{sec:conclusion}

Agreement protocols for nonlinear multi-agent dynamics over cooperative or cooperative--antagonistic networks were investigated. A class of nonlinear control laws were introduced based on relaxed convexity conditions. The price to pay was that each agent must have access to the orientations of a shared coordinate system, similar to a magnetic compass. Each agent specified a local supporting hyperrectangle with the help of the shared reference directions and the relative state measurements, and then a strict tangent cone was determined. Under mild conditions on the nonlinear dynamics and the interaction graph, we proved that for cooperative networks, exponential state agreement is achieved if and only if the interaction graph is uniformly jointly quasi-strongly connected. For cooperative--antagonistic networks, the componentwise absolute values of all the agents converge to the same values if the time-varying interaction graph is uniformly jointly strongly connected. The results generalize existing studies on agreement seeking of multi-agent systems. Future works include higher-order agent dynamics, convergence conditions for bipartite agreement, and the study on the case of mismatched shared reference directions.

\bibliographystyle{plain}
\bibliography{refs}

\noindent{\bf \large Appendices}

Note that a function $h(\cdot)$ is called Lipschitz continuous on a set $\mathcal{U}$ if there exists a constant $L_{\mathcal{U}}$ such that $\|h(a)-h(b)\|\leq L_{\mathcal{U}} \|a-b\|$ for all $a,b\in \mathcal{U}$.
\begin{lem}\label{lem:Lipschitz1-1}
Suppose Assumption \ref{assm1} holds, i.e., $f_p,~p\in\mathfrak{P}$, is uniformly locally Lipschitz. Assume that there exists a point $z_0\in \mathbb{R}^{dn}$ such that $\sup_{p\in \mathfrak{P}} f_p (z_0)$ (or $\inf_{p\in \mathfrak{P}} f_p (z_0)$) is finite. Then $g(x):=\sup_{p\in \mathfrak{P}} f_p (x)$ (or $\inf_{p\in \mathfrak{P}} f_p (x)$) is well defined and is Lipschitz continuous on every compact set $\mathcal{U}$.
\end{lem}

{\it Proof.}
Let $\mathcal{U}$ be a given compact set. Define $\mathcal{U}_{z_0}= {\rm co}(\{z_0\}\cup\mathcal{U})$. Based on Theorem 1.14 of \cite{Markley_Book}, a locally Lipschitz function is Lipschitz continuous on every compact subset. Plugging in the fact that $f_p, p\in\mathfrak{P}$ is uniformly locally Lipschitz, there is $L_{\mathcal{U}_{z_0}}>0$ such that $\|f_p(a)-f_p(b)\|\leq L_{\mathcal{U}_{z_0}} \|a-b\|$ for all $a,b\in \mathcal{U}_{z_0}$ and $p\in \mathfrak{P}$. It becomes straightforward that $g(x)$ is finite at every point in $\mathcal{U}_{z_0}$ and $L_{\mathcal{U}_{z_0}}$ is a Lipschitz constant of $g$ on $\mathcal{U}_{z_0}$. Therefore, the lemma holds.
\hfill$\square$

The following lemma is originally from \cite{LinZhiyun_PhD} and restated here.
\begin{lem}\label{lem:Lipschitz1-2}
Suppose that $f(x_1,y):\mathbb{R}\times\mathcal{M}\rightarrow \mathbb{R}$ is locally Lipschitz with respect to $[x_1,y]\T$, where $\mathcal{M}\subset \mathbb{R}^q$ is compact. Then $g(x_1)=\max_{y\in \mathcal{M}}f(x_1,y)$ is locally Lipschitz.
\end{lem}

\begin{lem}\label{lem:Lipschitz2-1}
Suppose that $f(x_1,y):\mathcal{M}_1\times\mathcal{M}\rightarrow \mathbb{R}$ is locally Lipschitz with respect to $[x_1,y]\T$, where $\mathcal{M}_1\subset \mathbb{R}$ and $\mathcal{M}\subset \mathbb{R}^q$ are compact. Then $g:\mathcal{M}_1\rightarrow \mathbb{R}$, $x_1\mapsto\max_{y\in \mathcal{M}}x_1f(x_1,y)$ is Lipschitz continuous with respect to $x_1$ on $\mathcal{M}_1$.
\end{lem}

{\it Proof.}
Because $f(x_1,y)$ is locally Lipschitz with respect to $[x_1,y]\T$ and $\mathcal{M}_1$ and $\mathcal{M}$ are compact, it follows that
$f(x_1,y)$ is Lipschitz continuous with respect to $[x_1,y]\T$. Therefore, there exists a constant $L$ such that
$
\|f(x_1,y)-f(\overline{x}_1,y)\|\leq L\|x_1-\overline{x}_1\|,~~\forall x_1,\overline{x}_1\in \mathcal{M}_1, ~\forall y\in \mathcal{M}.
$
Also, since $f(x_1,y)$ is continuous and the continuous function on the compact set is compact, there exist constants $L_x$ and $L_f$ such that $L_x=\max_{x_1\in \mathcal{M}_1}\|x_1\|$ and $L_f=\max_{x_1\in \mathcal{M}_1,y\in\mathcal{M}}\|f(x_1,y)\|$.

Let $y_x$ and $y_{\overline{x}}$ be the points satisfying $g(x_1)=\max_{y\in \mathcal{M}}\{x_1f(x_1,y)\}=x_1f(x_1,y_x)$ and $g(\overline{x}_1)=\max_{y\in \mathcal{M}}\{\overline{x}_1f(\overline{x}_1,y)\}=\overline{x}_1f(\overline{x}_1,y_{\overline{x}})$. It is trivial to show that $x_1f(x_1,y_x)\geq x_1f(x_1,y_{\overline{x}})$ and $\overline{x}_1f(\overline{x}_1,y_{\overline{x}})\geq \overline{x}_1f(\overline{x}_1,y_x)$.
Therefore, there exists $\widetilde{x}=(1-\lambda)x_1+\lambda\overline{x}_1$, where $0\leq \lambda\leq 1$ such that $\widetilde{x}f(\widetilde{x},y_{x})=\widetilde{x}f(\widetilde{x},y_{\overline{x}})$. Thus,
$
\|g(x_1)-g(\overline{x}_1)\|=~\|x_1f(x_1,y_x)-\widetilde{x}f(\widetilde{x},y_{x})
+\widetilde{x}f(\widetilde{x},y_{\overline{x}})-\overline{x}_1f(\overline{x}_1,y_{\overline{x}})\|
\leq~\|x_1f(x_1,y_x)-x_1f(\widetilde{x},y_{x})\|+\|x_1f(\widetilde{x},y_{x})-\widetilde{x}f(\widetilde{x},y_{x})\|
+\|\widetilde{x}f(\widetilde{x},y_{\overline{x}})-\widetilde{x}f(\overline{x}_1,y_{\overline{x}})\|
+\|\widetilde{x}f(\overline{x}_1,y_{\overline{x}})-\overline{x}_1f(\overline{x}_1,y_{\overline{x}})\|.
$
It then follows that
\begin{align}
\|g(x_1)-g(\overline{x}_1)\|
\leq&~ L\|x_1\|\|x_1-\widetilde{x}\|+\|f(\widetilde{x},y_{x})\|\|x_1-\widetilde{x}\|
+L\|\widetilde{x}\|\|\overline{x}_1-\widetilde{x}\|+\|f(\overline{x}_1,y_{\overline{x}})\|\|\overline{x}_1-\widetilde{x}\|
\notag\\ \leq&~(LL_x+L_f)\|x_1-\widetilde{x}\|+(LL_x+L_f)\|\overline{x}_1-\widetilde{x}\|
\notag\\~=&~(LL_x+L_f)\|x_1-\overline{x}_1\|.
\end{align}
Therefore, $g(x_1)$ is Lipschitz continuous with respect to $x_1$ on $\mathcal{M}_1$.
\hfill$\square$

\begin{lem}\label{lem:Lipschitz3}
Suppose that $f(x):\mathcal{M}\rightarrow \mathbb{R}$ is locally Lipschitz with respect to $x$, where $\mathcal{M}\subset \mathbb{R}^q$ is compact. Then $g(x):\mathcal{M}\rightarrow \mathbb{R}$, $x\mapsto x_1f(x)$ is Lipschitz continuous with respect to $x$ on $\mathcal{M}$, where $x_1$ denotes an element of $x$.
\end{lem}

{\it Proof.}
Because $f(x)$ is locally Lipschitz with respect to $x$ and $\mathcal{M}$ is compact, it follows that
$f(x)$ is Lipschitz continuous with respect to $x$. Therefore, there exists a constant $L$ such that
$
\|f(x)-f(\overline{x})\|\leq L\|x-\overline{x}\|,~~\forall x,\overline{x}\in \mathcal{M}.
$
Also, since $f(x)$ is continuous and the continuous function on the compact set is still compact, there exist constants $L_x$ and $L_f$ such that $L_x=\max_{x\in \mathcal{M}}\|x\|$ and $L_f=\max_{x\in\mathcal{M}}\|f(x)\|$.

It then follows that $\forall x,\overline{x}\in \mathcal{M}$,
$
\|g(x)-g(\overline{x})\|=~\|x_1f(x)-\overline{x}_1f(\overline{x})\|
\leq~LL_x\|x-\overline{x}\|+L_f\|x-\overline{x}\|
=~(LL_x+L_f)\|x-\overline{x}\|.
$
Therefore, $g(x)$ is Lipschitz continuous with respect to $x$ on $\mathcal{M}$.
\hfill$\square$

\end{document}